\documentclass[aps,pra,twocolumn,nofootinbib,maxnames=4]{revtex4-2}
\usepackage[colorlinks=true, allcolors=blue]{hyperref}
\usepackage[utf8]{inputenc}
\usepackage{graphicx}
\usepackage{multirow}
\usepackage{ulem}
\usepackage{amssymb}
\usepackage{amsmath}
\usepackage{microtype}
\usepackage{bm}
\usepackage{bbm}
\usepackage{enumitem}
\usepackage{bbold}
\usepackage{footnote}
\usepackage{mathrsfs}
\usepackage{revsymb}
\DeclareMathAlphabet{\mathcursive}{U}{rsfs}{m}{n}

\begin{document}
\title{Atomic parity violation in highly charged $^{40,48}$Ca and $^{208}$Pb ions}
	
\author{A.~V.~Viatkina$^{1,2}$, Ch.~G.~Mertens$^{1,3}$, B.~Ohayon$^{4}$, V.~A.~Yerokhin$^{5}$, A.~Surzhykov$^{1,2}$}

\affiliation{$^1$Technische Universität Braunschweig, 38106 Braunschweig, Germany}
\affiliation{$^2$Physikalisch-Technische Bundesanstalt, 38116 Braunschweig, Germany}
\affiliation{$^3$Rheinisch-Westfälische Technische Hochschule Aachen, 52062 Aachen, Germany}
\affiliation{$^4$Technion-Israel Institute of Technology, Haifa 3200003, Israel}
\affiliation{$^5$Max Planck Institute for Nuclear Physics, 69117 Heidelberg, Germany}
	
\begin{abstract}
We calculate parity-violation-induced E1 amplitudes for the $1s\rightarrow 2s$ and $1s^2 2s\rightarrow 1s^2 3s$ transitions in H- and Li-like ions of $^{40}$Ca, $^{48}$Ca, and $^{208}$Pb. In our analysis, we account for neutron skin effects and nuclear uncertainties for each nucleus. We consider the spin-independent weak-interaction contribution of the $Z^0$ boson described by standard model, as well as the effects of a hypothetical new $Z'$ boson of varying mass. We conclude that the neutron-skin corrections in the $^{40,48}$Ca isotope pair can mostly be neglected when considering $Z'$ boson effects, which is an advantage for the search for new parity-violating physics. On the other hand, both the neutron skin effect and the sensitivity to hypothetical $Z'$ interactions in $^{208}$Pb are shown to be significant.
\end{abstract}
	
\maketitle
	
\section{Introduction}
Atomic parity violation (APV) is one of the leading tools for testing the standard model (SM) and its possible extensions in the low-energy regime \cite{bouchiatParityViolationAtoms1997, gingesViolationsFundamentalSymmetries2004,dereviankoTheoreticalOverviewAtomic2007,robertsParityTimeReversalViolation2015}. Within the SM, the violation of parity symmetry occurs due to the weak interaction \cite{ParticleDataGroup:2024cfk}. Consequently, if the APV contribution is measured to be different from theoretical predictions based on the SM, it may indicate the existence of a new parity-violating interaction, for instance, one mediated by a new light $Z'$ boson \cite{safronovaSearchNewPhysics2018, dzubaProbingLowMassVector2017}. Alternatively, bounds can be placed on the parameters of such hypothetical couplings when the APV effects are found to agree with the SM.

To this day, the majority of APV measurements have been performed in neutral atoms~\cite{woodMeasurementParityNonconservation1997,antypasIsotopicVariationParity2019,barkovParityViolationAtomic1979}. However, given the prospect of storage-ring-based spectroscopy exemplified by Gamma Factory project at CERN \cite{krasnyGammaFactoryProposal2015}, there is a real possibility of probing APV in photoexcitations of highly charged ions (HCIs) \cite{richterParityViolationStudiesPartially2022, budkerAtomicPhysicsStudies2020}. In these proposed experiments, the ions moving with relativistic velocities in a storage ring are irradiated with a counter-propagating laser beam whose energy is increased in the ions' frame of reference due to the Doppler shift. In the Gamma Factory scenario, the Doppler-boosted laser photons' energies are expected to be up to 60~keV, which would facilitate spectroscopy of high-energy transitions from the ground state of many HCIs \cite{budkerAtomicPhysicsStudies2020}.

There are several advantages of APV experiments with HCIs over those in neutral atoms or multiply-charged ions: most notably, HCIs have a much simpler electronic structure, which is favorable for theoretical calculations, as well as a larger overlap of electronic wave functions with the nucleus that results in an enhancement of the APV effect \cite{budkerAtomicPhysicsStudies2020}. 

Apart from the new-physics searches, APV can provide insights into nuclear neutron distributions, which, in turn, help determine the parameters of the equation of state for neutron-rich matter \cite{thielNeutronSkinsAtomic2019}. These distributions are often characterized by the so-called \textit{neutron skin}, which is the difference between the root-mean-square (rms) radii of neutron and proton densities. While the proton distribution is relatively easy to measure in experiments relying on the electromagnetic interaction, the neutron distribution cannot be probed in the same straightforward way, since neutrons lack electric charge. Consequently, the neutron rms radii are poorly known in comparison to their proton counterparts. On the other hand, when $Z^0$-boson-mediated weak interaction is considered, electrons couple to neutrons much more strongly than they do to protons \cite{ParticleDataGroup:2024cfk}. Hence, the dominant part of the APV effect depends on the neutron density and as a result, the APV magnitude can be used to probe neutron distributions.

One can note that APV is not the only way to measure nuclear neutron density. Recently, neutron skins of $^{48}$Ca and $^{208}$Pb have been determined in parity-violating electron scattering experiments---in CREX \cite{crexcollaborationPrecisionDeterminationNeutral2022} and PREX-II \cite{prexcollaborationAccurateDeterminationNeutron2021} collaborations, respectively. In the present work, we use these results to calculate the APV amplitudes in H- and Li-like $^{40, 48}$Ca and $^{208}$Pb, taking into account the difference between proton and neutron rms radii. The significance of the pair $^{40}$Ca and $^{48}$Ca arises from the fact that both isotopes have relatively small neutron skins and almost equal charge radii. From the new-physics-search perspective,  these characteristics of $^{40, 48}$Ca nuclei help separate new parity-violating interactions coupling to protons from those coupling to neutrons.

The paper is organized as follows. In Sec.~\ref{sec:theory}, we examine the theory behind spin-independent standard-model (Sec.~\ref{sec:theory:SM_APV}) and hypothetical new-physics (Sec.~\ref{sec:theory:new_phys}) APV interactions. Sec.~\ref{sec:nsk_and_np} deals with the interplay between the neutron-skin and new-physics effects in APV by considering the relevant atomic matrix elements (Sec.~\ref{sec:nsk_and_np:ME}), their differences, and their ratios (Sec.~\ref{sec:nsk_and_np:isotop_diff}). We present our results in Sec.~\ref{sec:results} and discuss the nuclear parameters (Sec.~\ref{sec:results:nuclear_parameters}) and the calculation methods (Sec.~\ref{sec:results:methods}) that we use. Finally, we analyze our findings in Sec.~\ref{sec:results:analysis}. The summary is given in Sec.~\ref{sec:conclusion}.

In this work, most results are presented in atomic units. We write these explicitly, including all relevant constants, such as the reduced Planck constant $\hbar$, the speed of light $c$, and the fine structure constant $\alpha$. We denote the electron mass as $m$, the elementary charge as $e$, and the Bohr radius as $a_B$.

\section{Theory}\label{sec:theory}
\subsection{Standard Model APV}\label{sec:theory:SM_APV}
A major mechanism of atomic parity violation is the neutral-current interaction between electrons and nucleons, mediated by the $Z^0$ boson. For spinless nuclei and low energies, this interaction can be expressed through an effective four-fermion Hamiltonian density \cite{ParticleDataGroup:2024cfk}:

\begin{equation}
    \hat{H}_\mathrm{SM}=\frac{G_F}{\sqrt{2}}\sum_i C^{(\mathcal{N})}_{1i}\,\overline{\psi}^{(e)}\gamma_\mu\gamma^5 \psi^{(e)}\overline{\psi}^{(\mathcal{N})}_i \gamma^\mu \psi^{(\mathcal{N})}_i, \label{eq:4-fermion_H}
\end{equation}
where $G_F=1.166\times 10^{-5} (\hbar c)^3$ GeV$^{-2}$ is the Fermi constant, $\gamma^\mu$ are the Dirac gamma matrices, and $\gamma^5=-i\gamma^0\gamma^1\gamma^2\gamma^3$ is the `parity-violation' gamma matrix. In Eq.~\eqref{eq:4-fermion_H}, the sum runs over all nucleons; $\psi^{(e)}$ designates electron spinors and $\psi^{(\mathcal{N})}_i$ nucleon (proton or neutron) spinors, while the coefficients $C^{(\mathcal{N})}_{1i}$ characterize the interaction between the electron and the nucleon. This interaction is described as point-like since we are interested in APV processes whose characteristic energy (and momentum transfer) is much lower than the mass of the $Z^0$ boson, $M_Z c^2\approx 91.2$~GeV.

If we assume that the nucleons are non-relativistic, Eq.~\eqref{eq:4-fermion_H} simplifies to an effective spin-independent parity violation Hamiltonian:
\begin{equation}
    \hat{h}_\mathrm{SM}\left(\vec{r}\right)=\frac{G_F}{\sqrt{2}}\left[C^{(p)}_{1}Z\rho_p(\vec{r})+C^{(n)}_{1}N\rho_n(\vec{r})\right]\gamma^5\ ,\label{eq:h_SI_01QED}
\end{equation}
where $\rho_p(\vec{r})$ and $\rho_n(\vec{r})$ are the normalized proton and neutron densities, $Z$ is the nuclear charge number, and $N$ is the neutron number. Here, we distinguish between the coupling coefficients $C^{(p)}_{1}$ for protons and $C^{(n)}_{1}$ for neutrons. Their SM tree-level expressions \cite{ParticleDataGroup:2024cfk} are
\begin{align}
    C^{(p)}_{1}&=\frac{1}{2}\left(1-4\sin^2\theta_W\right)\approx \frac{1}{2}\times 0.045 \ ,\label{eq:C1p}\\
    C^{(n)}_{1}&=-\frac{1}{2}\ ,\label{eq:C1n}
\end{align}
where $\theta_W$ is the Weinberg angle and $\sin^2\theta_W=0.23873(5)$ for low energies \cite{ParticleDataGroup:2024cfk}. Including radiative corrections to the 0.1\% level of accuracy \cite{ParticleDataGroup:2024cfk,erlerWeakNeutralCurrent2013,blundenGammaZBoxCorrections2012} results in the following values for the coupling coefficients:
\begin{align}
    C^{(p)}_{1}&=\frac{1}{2}\times 0.071, \label{eq:C1p_rad}\\
    C^{(n)}_{1}&=-\frac{1}{2}\times 0.989. \label{eq:C1n_rad}
\end{align}
Note that since $|C^{(n)}_{1}|\gg |C^{(p)}_{1}|$, the dominant effect in spin-independent APV comes from the neutrons.

The Hamiltonian \eqref{eq:h_SI_01QED} depends on both proton $\rho_p(\vec{r})$ and neutron $\rho_n(\vec{r})$ nuclear densities. 
Assuming the simplified case $\rho_p\approx \rho_n\approx \rho$, one finds
\begin{equation}
    \hat{h}_\mathrm{SM}\left(\vec{r}\right)\approx\frac{G_F}{2\sqrt{2}}Q_W\rho(\vec{r})\gamma^5,\label{eq:h_SI_Qw}
\end{equation}
where the nuclear weak charge $Q_W$ is introduced. Based on Eqs.~(\ref{eq:C1p}) and (\ref{eq:C1n}), the SM tree-level expression for $Q_W$ is
\begin{equation}
    Q_W=-N+Z(1-4\sin^2\theta_W).
\end{equation}
When radiative corrections to $C^{(p)}_1$ and $C^{(n)}_1$ are included, as shown in Eqs. \eqref{eq:C1p_rad} and \eqref{eq:C1n_rad}, the weak charge is modified \cite{ParticleDataGroup:2024cfk,erlerWeakNeutralCurrent2013,blundenGammaZBoxCorrections2012}.
In the present work, however, our main focus is on the instances with $\rho_n\neq \rho_p$. In such a case, the spin-independent Hamiltonian reads as Eq.~\eqref{eq:h_SI_01QED}, 
where $C^{(p,n)}_{1}$ are described by Eqs. (\ref{eq:C1p_rad}) and (\ref{eq:C1n_rad}). In what follows, we will use this Hamiltonian for our calculations.

\subsection{New APV Forces}\label{sec:theory:new_phys}
In the previous section, we investigated parity violation as described by the standard model. Here, we consider a hypothetical $Z'$ boson of mass $m_{Z'}$ which, in addition to $Z^0$, couples electrons to nucleons. If we treat nucleons nonrelativistically, the presence of the $Z'$ boson will amount to a parity-violating Yukawa potential \cite{dzubaProbingLowMassVector2017}
\begin{equation}\label{eq:V_12_fermions}
V(\vec{r}-\vec{R})=\frac{g_{e \mathcal{N}}}{4\pi}\ \hbar c\ \frac{e^{-m_{Z'} c |\vec{r}-\vec{R}| / \hbar}}{|\vec{r}-\vec{R}|} \gamma^5\ ,
\end{equation}
where $|\vec{r}-\vec{R}|$ is the distance between the electron located at $\vec{r}$ and the nucleon located at $\vec{R}$, and the coupling constant $g_{e \mathcal{N}}$ quantifies the interaction strength. As before, we are interested only in the nuclear-spin-independent effect of the $Z'$ boson. Consequently, to obtain a Hamiltonian suitable for atomic calculations, we need to integrate Eq.~\eqref{eq:V_12_fermions} over the nucleon density:
\begin{equation}
	\hat{h}_{Z'}(\vec{r})=\hbar c\gamma^5 \frac{g_{e \mathcal{N}}}{4\pi} N_\mathcal{N}\int \rho_\mathcal{N}(\vec{R}) \frac{e^{-m_{Z'} c\left|\vec{r}-\vec{R}\right|/\hbar}}{|\vec{r}-\vec{R}|} d\vec{R}.\label{eq:V_integral}
\end{equation}
Here $N_\mathcal{N}=Z, N$ is the number of nucleons (protons or neutrons) and $\rho_\mathcal{N}(\vec{r})$ the normalized nucleon density. In the heavy-mass limit $m_{Z'}\rightarrow\infty$ and assuming the spherical distribution of nucleons $\rho_\mathcal{N}(\vec{r})=\rho_\mathcal{N}(r)$, Eq.~\eqref{eq:V_integral} simplifies to
\begin{equation}\label{eq:V_mZ_infinity}
	\hat{h}_{Z'}(r) = \frac{g_{e \mathcal{N}}}{m_{Z'}^2c^4}\,(\hbar c)^3 N_\mathcal{N}\rho_\mathcal{N}(\vec{r}) \gamma^5,
\end{equation}
which has the same structure as the Hamiltonian \eqref{eq:h_SI_01QED} above.

\section{Neutron Skin and New Physics}\label{sec:nsk_and_np}
Weak interaction \eqref{eq:h_SI_01QED} mixes opposite-parity levels in atomic systems. As a result, atomic transitions that are forbidden under parity selection rules become weakly allowed. A transition of this type, which is especially relevant to experiment \cite{woodMeasurementParityNonconservation1997,antypasIsotopicVariationParity2019,craikEntanglementProtocolMeasure2025}, is the PV-induced E1 transition between two levels of the same parity. Labeling these levels $|a\rangle$ and $|b\rangle$, one can obtain the amplitude of the photoexcitation $|a\rangle \rightarrow |b\rangle$ within second-order perturbation theory:
\begin{equation}\label{eq:E_PV}
    \mathcal{E}_\mathrm{PV}=\sum_k\left(  \frac{\langle b|\hat{D}_z|k\rangle \langle k|\hat{h}_\mathrm{SM}|a\rangle}{E_a-E_k}+ \frac{\langle b|\hat{h}_\mathrm{SM}|k\rangle \langle k|\hat{D}_z|a\rangle}{E_b-E_k} \right),
\end{equation}
where $E_a < E_b$ and the incident light is assumed to be linearly polarized in the $z$ direction. Moreover, in Eq.~\eqref{eq:E_PV}, $\hat{D}_z$ is the $z$-component of the electron dipole operator, and the sum runs over all intermediate atomic states $k$ with energies $E_k$. 

\begin{center}
		\begin{figure}[tb]
			\centering
			\includegraphics[width=\linewidth]{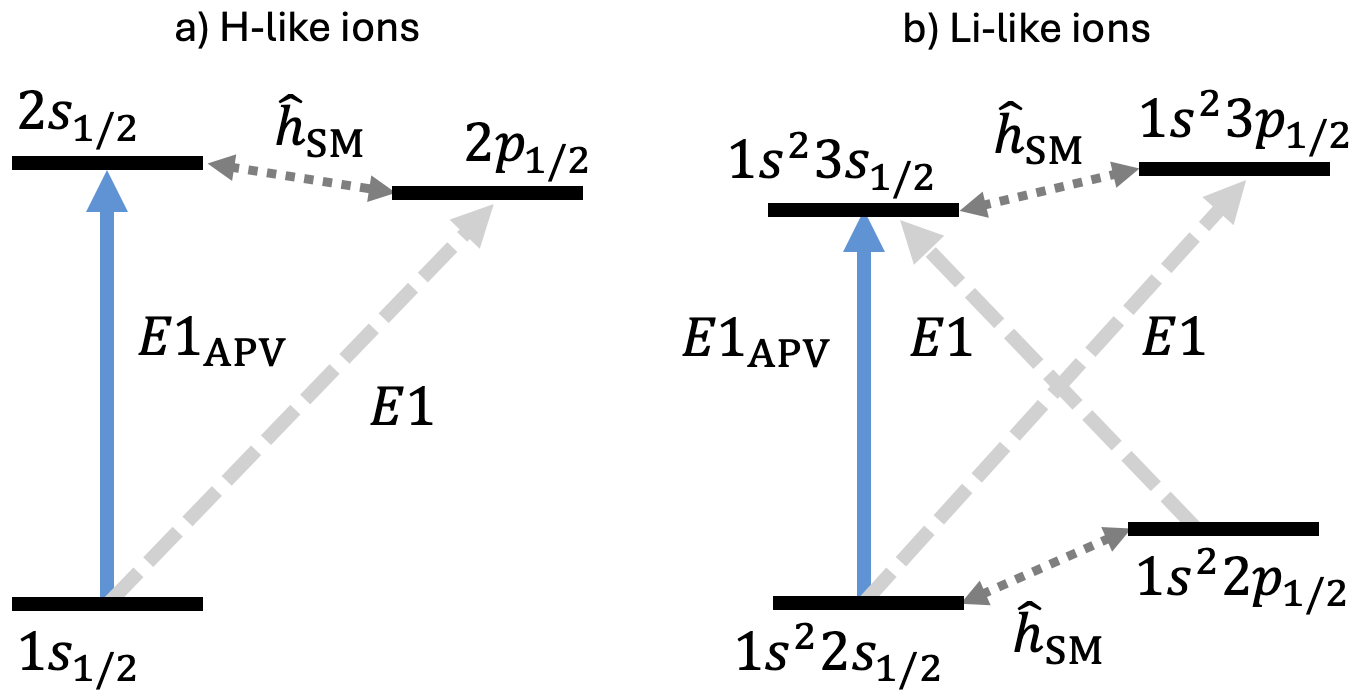}
			\caption{APV-induced E1 transition between the ground and the first excited $s$ states in a) hydrogenlike and b) lithiumlike ions. In each case, only the levels providing the dominant contributions to the sum in Eq.~\eqref{eq:E_PV} are shown. The PV-mixing introduced by the Hamiltonian $\hat{h}_\mathrm{SM}$ between the nearly-degenerate $s$ and $p_{1/2}$ states is depicted using dotted lines, the electric-dipole-allowed transitions are represented by dashed arrows.}
			\label{fig:transitions}
		\end{figure}
\end{center}

In the present work, we consider transitions between the ground and the first excited $s$-states in H-like and Li-like ions. In such a case, the sum in Eq.~\eqref{eq:E_PV} runs effectively over the $|k\rangle = |np_{1/2}\rangle$ intermediate states. The main contribution to this sum arises from just a few terms corresponding to the intermediate levels that are nearly degenerate with respect to either the initial $a$ or the final $b$ atomic state. More specifically, for the $1s\rightarrow 2s$ transition in H-like ions, there is a single dominant term with $|k\rangle = |2p_{1/2}\rangle$, and for the $1s^2 2s\rightarrow 1s^2 3s$ transition in Li-like ions, there are two such terms, with $|k\rangle = |1s^2 2p_{1/2}\rangle$ and with $|k\rangle = |1s^2 3p_{1/2}\rangle$, see Fig.~\ref{fig:transitions}.

\subsection{Matrix Elements}\label{sec:nsk_and_np:ME}

In order to illustrate the role of different contributions to atomic parity violation, let us consider the simplest case of the $1s\rightarrow 2s$ transition in a H-like ion. Additionally, let us limit ourselves to the dominant term of the sum in Eq.~\eqref{eq:E_PV}:
\begin{equation}\label{eq:E_PV_H-like_term}
	\mathcal{E}_\mathrm{PV}\approx \frac{\langle 2s|\hat{h}_\mathrm{SM}|2p_{1/2} \rangle \langle 2p_{1/2}|\hat{D}_z|1s\rangle}{E_{2s}-E_{2p_{1/2}}}\ .
\end{equation}
The matrix element of the operator $\hat{h}_\mathrm{SM}$ reads
\begin{align}\label{eq:ME_0}
	\mathcal{M} = \langle 2s|\hat{h}_\mathrm{SM}&|2p_{1/2} \rangle\nonumber \\ &=\int \psi_{2s}^\dagger(\vec{r}) \hat{h}_\mathrm{SM}(\vec{r}) \psi_{2p_{1/2}}(\vec{r})d\vec{r} ,
\end{align}
where $\psi_{2s}(\vec{r})$ and $\psi_{2p_{1/2}}(\vec{r})$ are the relativistic electron wavefunctions. By employing the explicit form of the operator $\hat{h}_\mathrm{SM}$ [see Eq.~\eqref{eq:h_SI_01QED}], we obtain
\begin{align}\label{eq:ME_0}
	\mathcal{M} = 
	\frac{G_F}{\sqrt{2}} &\left( C^{(p)}_{1}Z\int \rho_p(\vec{r}) \psi_{2s}^\dagger(\vec{r})\gamma^5 \psi_{2p_{1/2}}(\vec{r})d\vec{r}\right. + \nonumber\\ 
	&\left. C^{(n)}_{1}N\int \rho_n(\vec{r}) \psi_{2s}(\vec{r})^\dagger\gamma^5 \psi_{2p_{1/2}}(\vec{r})d\vec{r}\right).
\end{align}
Assuming that both proton $\rho_p$ and neutron $\rho_n$ densities are spherically symmetric and normalized to unity, $4\pi \int \rho_{p,n}(r)r^2 dr =1$, we can write 
\begin{align}
	&\mathcal{M} = 
	\frac{G_F}{\sqrt{2}}\times\nonumber\\ 
	&\left( C^{(p)}_{1}Z\int dr r^2 \rho_p(r) \int\!\int d\phi d\theta \sin\theta \psi_{2s}^\dagger(\vec{r})\gamma^5 \psi_{2p_{1/2}}(\vec{r}) \right. + \nonumber\\ 
	&\left. C^{(n)}_{1}N\int dr r^2 \rho_n(r) \int\!\int d\phi d\theta \sin\theta \psi_{2s}^\dagger(\vec{r})\gamma^5 \psi_{2p_{1/2}}(\vec{r})\right).\label{eq:M_integral_long}
\end{align}
One can see that both terms have the same angular integral, which can be evaluated using the standard representation of Dirac wavefunctions:
\begin{equation}
\psi_{n\kappa m}\left(\vec{r}\right) = \frac{1}{r} \begin{pmatrix}
f_{n\kappa}(r)\Omega_{\kappa m}(\theta,\phi)\\
i g_{n\kappa}(r)\Omega_{-\kappa m}(\theta,\phi)
\end{pmatrix},
\end{equation}
where $f_{n\kappa}$ and $g_{n\kappa}$ are the radial components and $\Omega_{\kappa m}$ are spherical spinors; $n$ is the principal quantum number, $\kappa$ the Dirac quantum number, and $m$ the projection of the angular momentum. Making use of this expression, we obtain
\begin{align}
	\int\!\int d\phi d\theta &\sin\theta \psi_{2s}^\dagger(\vec{r})\gamma^5 \psi_{2p_{1/2}}(\vec{r})=\nonumber \\
	&\frac{i}{r^2}\left[ f_{2s}(r)g_{2p_{1/2}}(r)-g_{2s}(r)f_{2p_{1/2}}(r) \right].\label{eq:radial_int_0}
\end{align}
To illustrate the effect of the neutron skin on the APV, it is instructive to rewrite Eq.~\eqref{eq:radial_int_0} as
\begin{align}
	\frac{i}{r^2}\left[ f_{2s}(r)g_{2p_{1/2}}(r)-g_{2s}(r)f_{2p_{1/2}}(r) \right]&\equiv \nonumber \\
	\frac{i}{r^2}&\mathcal{N}_\psi f_{\psi}(r),\label{eq:radial_int_1}
\end{align}
where the normalization factor $\mathcal{N}_\psi$ is chosen in such a manner that $f_{\psi}(0)=1$. One can present $\mathcal{N}_\psi$ as a product
\begin{equation}
	\mathcal{N}_\psi=\mathcal{A}\mathcal{P},\label{eq:N_psi}
\end{equation}
where $\mathcal{A}$ depends exclusively on the electronic configuration, while $\mathcal{P}$ includes nuclear parameters and can be approximated \cite{viatkinaDependenceAtomicParityviolation2019} as
\begin{equation}\label{eq:P_norm}
	\mathcal{P}\approx\left(2 Z r_p/a_0\right)^{2\gamma -2} ,
\end{equation}
where $a_0$ is the Bohr radius, $r_p$ is the nuclear rms charge radius, and $\gamma=\sqrt{1-Z^2\alpha^2}$.
By inserting Eqs.~(\ref{eq:radial_int_1})--(\ref{eq:N_psi}) into Eq.~\eqref{eq:ME_0}, we can express the matrix element of the operator $\hat{h}_\mathrm{SM}$ in the form 
\begin{equation}\label{eq:ME}
	\mathcal{M} = i \frac{G_F}{\sqrt{2}}\mathcal{A}\mathcal{P}\left(C_{1p} Z q_p + C_{1n} N q_n \right),
\end{equation}
where
\begin{equation}\label{eq:q_pn}
	q_{p,n} = \int f_\psi(r)\rho_{p,n}(r) dr,
\end{equation}
is the integral of the normalized function $f_\psi(r)$ over nuclear density. One can further simplify the matrix element \eqref{eq:ME} as
\begin{equation}
	\mathcal{M} = i \frac{G_F}{2\sqrt{2}}\mathcal{A}\mathcal{P}\tilde{Q}_W ,
\end{equation}
where we introduced an \textit{effective weak charge}
\begin{equation}
	\tilde{Q}_W = 2 C_{1p} Z q_p + 2 C_{1n} N q_n .
\end{equation}
Note that for the case $q_p=q_n\equiv q$ this effective parameter reads as $\tilde{Q}_W = q\cdot Q_W$, where $Q_W$ is the conventional weak charge.

Until now, we have discussed the evaluation of the matrix element \eqref{eq:ME_0} within the framework of the standard model. A hypothetical new-physics interaction mediated by a $Z'$ boson, described by the Hamiltonian \eqref{eq:V_integral}, would modify the matrix element as
\begin{gather}
	\mathcal{M} = \langle 2s|\hat{h}_\mathrm{SM}+\hat{h}_{Z'}|2p_{1/2}\rangle =\nonumber\\ 
	i\frac{G_F}{2\sqrt{2}}\mathcal{A}\mathcal{P}\left[Z\left( 2C_{1p}q_p + \Delta\tilde{Q}_p\right) + N\left(2C_{1n}q_n +\Delta\tilde{Q}_n\right) \right],
\end{gather} 
where we included small hypothetical new-physics corrections $\Delta\tilde{Q}_p$ and $\Delta\tilde{Q}_n$ to the proton and neutron contributions to the effective weak charge. These corrections can be derived in a similar way to $\tilde{Q}_W$ but they imply the interaction operator \eqref{eq:V_integral}.

\subsection{Isotopic Difference and Ratio for $^{40,48}$Ca}\label{sec:nsk_and_np:isotop_diff}
It is instructive to study how $\mathcal{E}_\mathrm{PV}$ changes between two isotopes of the same atom. Again, let us consider the simplest case of the dominant term in the $1s\rightarrow 2s$ transition of a H-like ion, see Eq.~\eqref{eq:E_PV_H-like_term}. Moreover, let us also neglect the isotope shift in the energy difference $E_{2s}-E_{2p_{1/2}}$ and the isotopic correction to the dipole matrix element $\langle 2p_{1/2}|\hat{D}_z| 1s\rangle$, which leaves only the `parity-violating' matrix element $\mathcal{M}$ to be investigated. To quantify the isotopic change in $\mathcal{M}$, let us consider the difference $\mathcal{M}'-\mathcal{M}$ and the ratio $\mathcal{M}'/\mathcal{M}$ of the PV matrix elements for two different isotopes, $A$ and $A'$. We start with the difference:
\begin{align}\label{eq:diff_general}
	&\mathcal{M}'-\mathcal{M}=i \frac{G_F}{2\sqrt{2}}\mathcal{A} \nonumber\\ & \times \left\{\mathcal{P}'\left[ Z \left(2C_{1p} q'_p + \Delta\tilde{Q}'_p\right) + N' \left(2C_{1n} q'_n + \Delta\tilde{Q}'_n\right) \right]\right. \nonumber \\ &\left. - \mathcal{P}\left[ Z \left(2C_{1p} q_p + \Delta\tilde{Q}_p\right) + N \left(2C_{1n} q_n + \Delta\tilde{Q}_n\right) \right]\right\}.
\end{align}
This expression can be simplified for the particular case of the Ca isotope pair ($A=40$, $A'=48$) by virtue of two observations. First, as will be shown in Sec.~\ref{sec:results:nuclear_parameters}, the proton charge distributions in $^{40}$Ca and $^{48}$Ca isotopes are very similar to each other, leading to $r'_p\approx r_p$, $\mathcal{P}'\approx \mathcal{P}$, $q'_p\approx q_p$, and $\Delta\tilde{Q}'_p\approx \Delta\tilde{Q}_p$. Hence, the `proton' terms in Eq.~\eqref{eq:diff_general} cancel out, and we can rewrite the difference as 
\begin{align}
	&\mathcal{M}'-\mathcal{M}=i \frac{G_F}{2\sqrt{2}}\mathcal{A}\mathcal{P}\nonumber \\& \times\left[N'\left(2C_{1n} q'_n + \Delta\tilde{Q}'_n\right)-N\left(2C_{1n} q_n + \Delta\tilde{Q}_n\right)  \right].
\end{align}
Second, the proton and neutron densities in the $^{40}$Ca isotope are also very similar (see Ref.~\cite{emrichRadialDistributionNucleons1983} and the discussion in Sec.~\ref{sec:results:nuclear_parameters} below), since $N=Z=20$, which leads to $q_p\approx q_n$. To proceed further, let us introduce a neutron skin contribution to the integral $q_n$:

\begin{equation}
	\delta q'_{n,\mathrm{nsk}} \equiv q'_n - q'_p \approx q'_n - q_n \label{eq:dqnsk},
\end{equation}
where we used the fact that $q'_p\approx q_p\approx q_n$ in the $^{40,48}$Ca isotope pair. On the other hand, it is reasonable to expect that the new parity-violating neutron coupling is very weak, so that we can neglect its neutron skin contribution
\begin{equation}
		\delta \tilde{Q}'_{n,\mathrm{nsk}} \equiv \Delta\tilde{Q}'_n - \Delta\tilde{Q}'_p \approx \Delta\tilde{Q}'_n - \Delta\tilde{Q}_n \approx 0 , \label{eq:DQnsk}
\end{equation}
since $\delta \tilde{Q}'_{n,\mathrm{nsk}}\ll \delta q'_{n,\mathrm{nsk}}$. By employing this fact and Eq.~\eqref{eq:dqnsk}, we finally obtain

\begin{align}
	\mathcal{M}'&-\mathcal{M}=i \frac{G_F}{2\sqrt{2}}\mathcal{A}\mathcal{P}N'\nonumber\\ 
	&\times\left[ \frac{\Delta N}{N'} \left( 2C_{1n}q_n+\Delta\tilde{Q}_n \right)+ 2C_{1n}\delta q'_{n,\mathrm{nsk}}\right]\ , \label{eq:Ca_M_difference}
\end{align}
where $\Delta N=N'-N$. We see that the difference of the PV matrix elements in the $^{40,48}$Ca isotope pair contains no terms arising from proton new physics couplings, and that the neutron new physics contribution $\Delta\tilde{Q}_W$ is somewhat suppressed by the factor $\Delta N/N'$. However, for the $^{40,48}$Ca isotope pair, $\Delta N$ is relatively large and $\Delta N/N'\approx 0.3$. In this way, the large $\Delta N$ difference in $^{40,48}$Ca provides an advantage for the search for new physics. 

Now let us turn to the ratio of the dominant PV matrix elements. For a H-like ion, this ratio reads as:
\begin{align}\label{eq:M_Ratio}
	&\frac{\mathcal{M}'}{\mathcal{M}}=\frac{\mathcal{P}'}{\mathcal{P}}\nonumber \\ &\times \frac{Z \left(2C_{1p} q'_p + \Delta\tilde{Q}'_p\right) + N' \left(2C_{1n} q'_n + \Delta\tilde{Q}'_n\right)}{Z \left(2C_{1p} q_p + \Delta\tilde{Q}_p\right) + N \left(2C_{1n} q_n + \Delta\tilde{Q}_n\right)} .
\end{align}
The correction to $\Delta\tilde{Q}_p$ due to the variation of the nuclear charge radius between two isotopes is very small; hence, one can assume that $\Delta \tilde{Q}'_p \approx \Delta \tilde{Q}_p$. Additionally, $|C_{1p}|\ll |C_{1n}|\approx 1/2$ according to Eqs.~\eqref{eq:C1p_rad}--\eqref{eq:C1n_rad}. Thus,  we can obtain a rough estimate
\begin{equation}
\frac{\mathcal{M}'}{\mathcal{M}}=\frac{\mathcal{P}'\left(-N'+\Delta \tilde{Q}'_W\right)}{\mathcal{P}\left(-N+\Delta \tilde{Q}_W\right)}\approx\frac{\mathcal{P'}N'}{\mathcal{P}N}\left[1+\frac{Z\Delta N}{NN'}\Delta \tilde{Q}_p\right],\label{eq:ratio_newphys}
\end{equation}
which shows that the ratios of PV matrix elements for different isotopes are primarily sensitive to the new proton couplings $\Delta \tilde{Q}_p$. 

While \eqref{eq:ratio_newphys} presents an estimate of a ME ratio for arbitrary isotopes, more detailed analysis of the general Eq.~\eqref{eq:M_Ratio} can be performed for the $^{40,48}$Ca pair. For this particular case, as noted above, $r'_p\approx r_p$, $\mathcal{P}'\approx \mathcal{P}$, $q'_p\approx q_p$, $\Delta\tilde{Q}'_p\approx \Delta\tilde{Q}_p$, and $q_p\approx q_n$. By making use of these approximations and Eqs.~(\ref{eq:dqnsk}) and (\ref{eq:DQnsk}), we obtain the matrix element ratio:
\begin{align}
\frac{\mathcal{M}'}{\mathcal{M}}\approx &\frac{N'}{N}\left(1+\frac{\delta q'_{n,\mathrm{nsk}}}{q_n}\right)\nonumber\\
	&\times\left[1-\frac{Z\Delta N}{N'N}\frac{2C_{1n}}{q_n}\left(2C_{1p}q_p+\Delta \tilde{Q}_p \right)\right],\label{eq:Ca_M_ratio}
\end{align}
where the new physics contribution is again suppressed by $\Delta N/N'\approx 0.3$. 

As seen from Eqs.~\eqref{eq:Ca_M_difference} and \eqref{eq:Ca_M_ratio}, the neutron skin $\delta q'_{n,\mathrm{nsk}}$ and the new physics $\Delta\tilde{Q}_{n,p}$ corrections enter the matrix element difference and ratio on an equal footing. Therefore, in order to investigate the new-physics PV effects, knowledge of the neutron skin corrections is required. In what follows, we will calculate those corrections for the $^{40,48}$Ca isotope pair. Moreover, even though it is not the main focus of our study, we will present the neutron skin and new physics corrections for $^{208}$Pb, since its neutron skin, similarly to $^{48}$Ca, has been measured \cite{prexcollaborationAccurateDeterminationNeutron2021}.

\section{Results}\label{sec:results}
\subsection{Nuclear Parameters}\label{sec:results:nuclear_parameters}
In our calculations, we use a normalized spherically symmetric Fermi distribution for nuclear charge densities
\begin{equation}\label{eq:Fermi_distribution}
\rho_{p}(r)=\frac{\rho_0}{1+\exp(\frac{r-c}{z})}\ ,
\end{equation}
where the normalization $\rho_0$ is chosen in such a way that $4\pi \int \rho_{p}(r)r^2 dr =1$. 
The values of the radius parameter $c$ and the diffuseness $z$ for $^{40,48}$Ca and $^{208}$Pb nuclei are displayed in Tab.~\ref{tab:nuc_param}. They are based on a combined analysis of electron scattering and muonic spectroscopy data. 

In our analysis, we assume that electron scattering experiments are best for measuring ratios of moments (mostly affected by the nuclear shape), while muonic atom x-ray spectroscopy is best for measuring a certain integral quantity called the Barrett moment.
For each of the three nuclei that we consider, we adjust the $c,z$ parameters until both the relevant Barrett moment (from muonic atoms) and the measured ratio of the fourth and second moments $\langle r^4 \rangle/\langle r^2 \rangle$, where $\langle r^n \rangle = \int r^n \rho_{p}(r) r^2 dr$, are reproduced~\cite{yerokhin2025model}.

The reported experimental uncertainty in scattering experiments is often small compared with the variation when considering different datasets and analysis methods~\cite{OHAYON2025101732}. In light of this, we consider two different parametrizations of the charge distribution for each isotope, with their differences approximating the uncertainty in the nuclear model.


\begin{table}[tb]
\caption{Fermi distribution parameters---namely, half-density radius $c$ and diffuseness $z$ [see Eq.~\eqref{eq:Fermi_distribution}], as well as the root-mean-square radius $r_\mathrm{rms}$, for the nuclear charge densities of $^{40,48}$Ca and $^{208}$Pb. They are chosen to reproduce the Barret moments given in~\cite{fricke2004nuclear} and the ratio of 4th to 2nd moment from two different methods.
For $^{40}$Ca, we either use the sum-of-Gaussians (SOG) or Fourier-Bessel (FB) parametrizations.
For $^{48}$Ca, we consider the SOG parametrization for $^{40}$Ca and the isotopic difference from either the upper or the lower curve in Fig.~3 of Ref.~\cite{emrichRadialDistributionNucleons1983}. 
For $^{208}$Pb, we consider either the parametrization in Ref.~\cite{devriesNuclearChargedensitydistributionParameters1987} or that in \cite{FRICKE1995177}, respectively.
\label{tab:nuc_param}
}

\begin{ruledtabular}
\begin{tabular}{l|llll}
Isotope       &  $c$, fm	    &  $z$, fm 	 &  $r_\mathrm{rms}$, fm 	&   Method \\
\hline
$^{40}$Ca	   &  $3.629$	&  $0.552$	 &	$3.480$ & SOG \cite{sickChargeDensity40Ca1979, devriesNuclearChargedensitydistributionParameters1987} \\ 
			   &	  $3.803$  	&  $0.493$	 &	$3.476$	& FB \cite{devriesNuclearChargedensitydistributionParameters1987} \\ 
\hline
$^{48}$Ca 	   &  $3.749$	&  $0.515$	 &  $3.478$ & SOG \cite{sickChargeDensity40Ca1979},  Ref.~\cite{emrichRadialDistributionNucleons1983} \\ 
			   &  $3.739$	&  $0.518$	 &  $3.478$ & SOG \cite{sickChargeDensity40Ca1979},  Ref.~\cite{emrichRadialDistributionNucleons1983} \\ 
\hline
$^{208}$Pb	   &  $6.663$   &  $0.513$  &  $5.502$  & \cite{devriesNuclearChargedensitydistributionParameters1987, EUTENEUER1978452} \\ 
			   &  $6.668$   &  $0.509$   &  $5.501$ & \cite{FRICKE1995177}  
\end{tabular}
\end{ruledtabular}
\end{table}

In order to model neutron densities in $^{40, 48}$Ca and $^{208}$Pb, we use the Fermi distribution \eqref{eq:Fermi_distribution} as well. For each nucleus, we take the parameters from Tab.~\ref{tab:nuc_param} as a starting point and, while keeping diffuseness $z$ intact, change the radius $c$ so that the difference between the root-mean-square radii $r_\mathrm{rms}=\sqrt{\langle r^2\rangle}$ of neutron and proton distributions $\Delta r_\mathrm{rms}= r_{n,\mathrm{rms}}-r_{p,\mathrm{rms}}$ correspond to the experimentally known neutron skins. For $^{40}$Ca, for example, we assume a small negative neutron skin
\begin{equation}
	\Delta r_\mathrm{rms}\left(^{40}\mathrm{Ca}\right)=-0.01(1)\, \mathrm{fm},\label{eq:Ca_nsk}
\end{equation}
based on the measurement in Ref.~\cite{emrichRadialDistributionNucleons1983} and an error estimate of 100\%. Neutron skins for $^{48}$Ca and $^{208}$Pb come from CREX  \cite{crexcollaborationPrecisionDeterminationNeutral2022} and PREX-II \cite{prexcollaborationAccurateDeterminationNeutron2021} experiments, respectively:\begin{align}
	\Delta r_\mathrm{rms}\left(^{48}\mathrm{Ca}\right)&=0.121(35)\, \mathrm{fm}, \label{eq:Ca48_nsk}\\
	\Delta r_\mathrm{rms}\left(^{208}\mathrm{Pb}\right)&=0.283(71)\, \mathrm{fm}, \label{eq:Pb_nsk}
\end{align}
where experimental and model errors were combined in quadrature.

\subsection{Numerical Calculations}\label{sec:results:methods}
In Sec.~\ref{sec:nsk_and_np}, we investigated the interplay between the neutron-skin and new-physics contributions for the model case of the $1s\rightarrow 2s$ transition in H-like ions and by restricting the intermediate-state summation in Eq.~\eqref{eq:E_PV} to a single dominant term. In what follows, we present numerical results for the general case: we calculate parity-violating amplitudes $\mathcal{E}_\mathrm{PV}$ \eqref{eq:E_PV} for H- and Li-like ions by employing summation over intermediate states. The calculations consist of two tasks: finding the matrix elements in the numerator and assessing the energy differences in the denominator. Let us start with the latter. Energy levels of H-like ions with principal quantum numbers $n=1, 2$ are tabulated in Ref.~\cite{yerokhinLambShiftStates2015}, where \textit{ab initio} QED calculations performed to all orders in $\alpha Z$ are presented. 
For levels with $n\geq 3$ we take the eigensolutions of Dirac equation with finite nuclear size, which are found using the \texttt{qm-dish} package \cite{mertensNumericalSolutionDirac} and where QED corrections are neglected. This approximation is justified because intermediate summation terms with $n\geq 3$ provide only a minor contribution to $\mathcal{E}_\mathrm{PV}$.

On the other hand, in Li-like ions the few first terms of the sum in Eq.~\eqref{eq:E_PV} have comparable magnitudes. Hence, we use the CI-QEDMOD approach described in Ref.~\cite{yerokhinEnergyLevelsCoreExcited2018} to calculate the energy levels $1s^2nl$ in Li-like $^{40}$Ca and $^{208}$Pb for $n\leq 6$, and \texttt{GRASP2K} \cite{jonssonGrasp2KRelativisticAtomic2007} for the levels with $n=7,8$. 
Our results for Li-like energies are summarized in Tab.~\ref{tab:Li_energies}. The energy isotope shifts in H- and Li-like ions were found to be smaller than the numerical uncertainty; hence, we neglect them in our calculations.

\begin{table}[tb]
\caption{Energy levels $1s^2nl$ of Li-like $^{40}$Ca and $^{208}$Pb.
Levels with $n\leq 6$ are calculated using CI-QEDMOD method~\cite{yerokhinEnergyLevelsCoreExcited2018}, including numerical uncertainties and isotope shifts in $^{48}$Ca with respect to $^{40}$Ca. Energies for $n= 7,8$ are found using \texttt{GRASP2K} \cite{jonssonGrasp2KRelativisticAtomic2007}; we estimate the errors for $n= 7,8$ in $^{40}$Ca by comparing our results to NIST data \cite{nist}, and in $^{208}$Pb by comparing to an \texttt{AMBiT} \cite{kahlAMBiTProgrammeHighprecision2019} calculation. \label{tab:Li_energies}}
\begin{ruledtabular}
\begin{tabular}{l|l|l|l}
Level  &$E_n\left(^{40}\mathrm{Ca}\right)$, eV & $\Delta E_n\left(^{40,48}\mathrm{Ca}\right)$, eV & $E_n\left(\mathrm{Pb}\right)$, eV \\
\hline
$1s^2 2s$	   &$0$          & $0$			&$0$	\\
$1s^2 2p$      &$35.961(1)$  & $0.0017(4)$	&$230.65(5)$	\\
$1s^2 3s$      &$651.828(2)$ & $0.0015(4)$	&$14179.98(8)$	\\
$1s^2 3p$      &$661.783(2)$ & $0.0019(4)$	&$14241.51(8)$	\\
$1s^2 4p$      &$879.564(2)$ & $0.0022(4)$	&$19037.23(8)$	\\
$1s^2 5p$	   &$980.2(1)$	 & $0.0023(4)$	&$21215(2)$	\\
$1s^2 6p$	   &$1035.7(1)$	 & $0.0024(4)$	&$22439(2)$	\\
$1s^2 7p$	   &$1067.1(2)$	 & $0.0025(4)$	&$23068(64)$	\\
$1s^2 8p$	   &$1088.3(2)$	 & $0.0025(4)$	&$23514(64)$	
\end{tabular}
\end{ruledtabular}
\end{table}

Now let us turn to the matrix elements of the $z$-component of electronic dipole operator $\hat{D}_z$ and of the parity violating Hamiltonians $\hat{h}_{SM}$ and $\hat{h}_{Z'}$ in Eq.~\eqref{eq:E_PV}. Using standard angular momentum algebra, one can split the relevant matrix elements into angular and radial parts. To calculate the radial integrals for H-like ions, we generate relativistic Dirac wavefunctions in a finite-size-nucleus potential using the \texttt{qm-dish} package \cite{mertensNumericalSolutionDirac}. For Li-like ions, we employ Dirac-Fock radial wavefunctions obtained from \texttt{GRASP2K} \cite{jonssonGrasp2KRelativisticAtomic2007}. For both cases, we use Fermi nuclear charge densities with the parameters from Tab.~\ref{tab:nuc_param}. Moreover, a fully relativistic formula presented in Ref.~\cite{kozlovSelfenergyCorrectionE12025} for the matrix elements of $\hat{D}_z$ is used, since the transition frequencies in H- and Li-like ions are high.

In order to calculate parity violating operators $\hat{h}_\mathrm{SM}$ and $\hat{h}_{Z'}$, one needs to know the proton and neutron nuclear densities, see Eqs.~\eqref{eq:h_SI_01QED} and \eqref{eq:V_integral}, which are described by Fermi distributions~\eqref{eq:Fermi_distribution}. For protons, the distribution parameters are listed in Tab.~\ref{tab:nuc_param}, while for neutrons they are derived from the neutron skin data, as discussed in Sec.~\ref{sec:results:nuclear_parameters}.

The dominant uncertainty for the PV matrix elements $\langle \psi_1|\hat{h}_\mathrm{SM}|\psi_2 \rangle$ and $\langle \psi_1|\hat{h}_\mathrm{Z'}|\psi_2 \rangle$ arises from the uncertainties of neutron skin and nuclear charge distribution. 
We estimated these uncertainties by calculating the matrix elements for two sets of parameters for each isotope, Tab.~\ref{tab:nuc_param}, and with different neutron skins within the error bars, see Eqs.~\eqref{eq:Ca_nsk} and \eqref{eq:Pb_nsk}. Based on these calculations, the matrix element $\langle \psi_1|\hat{h}_\mathrm{Z'}|\psi_2 \rangle$ for  lighter $Z'$ bosons turns out to be much less sensitive to the fine details of nuclear distributions than its standard-model counterpart $\langle \psi_1|\hat{h}_\mathrm{SM}|\psi_2 \rangle$. Therefore, we do not typically present the corresponding errors 
for the hypothetical bosons with masses $m_{Z'}\lesssim 10$~MeV.

\subsection{Analysis}\label{sec:results:analysis}
Having reviewed the calculations of atomic energy levels and matrix elements of the operators $\hat{D}_z$, $\hat{h}_\mathrm{SM}$, $\hat{h}_{Z'}$, we discuss the obtained results below. In particular, we move our focus onto the interplay between neutron skin and new physics. For the purposes of our discussion, we split the standard-model (SM) and new-physics (NP) PV matrix elements into neutron and proton parts. Following Eq.~\eqref{eq:h_SI_01QED}, the SM leading matrix element for H-like ions can be written as
\begin{equation}
	\mathcal{M}^\mathrm{(SM)}=C^{(p)}_1 Z \mathrm{m}^\mathrm{(SM)}_{p} + C^{(n)}_1 N \mathrm{m}^\mathrm{(SM)}_{n}\ ,\label{eq:M_SM_split}
\end{equation}
where
\begin{equation} 
\mathrm{m}^\mathrm{(SM)}_{p,n}=\langle 2s|\hat{h}_{\mathrm{SM},p,n}|2p_{1/2} \rangle ,\label{eq:me_SM}
\end{equation} 
are the matrix elements of the partial operator
\begin{equation}
	\hat{h}_{\mathrm{SM}, p,n}=\frac{G_F}{\sqrt{2}}\,\rho_{p,n}(\vec{r})\gamma^5 .\label{eq:h_SI_pn}
\end{equation}
By rewriting the matrix elements for all intermediate states in the form analogous to Eq.~\eqref{eq:M_SM_split}, we arrive at the PV amplitude:
\begin{equation}
	\mathcal{E}_\mathrm{PV}^{(\mathrm{SM})} = C^{(p)}_1 Z \varepsilon^\mathrm{(SM)}_{\mathrm{PV}, p} + C^{(n)}_1 N \varepsilon^\mathrm{(SM)}_{\mathrm{PV}, n}\ ,\label{eq:E_SM_split}
\end{equation}
where the partial amplitudes $\varepsilon^\mathrm{(SM)}_{\mathrm{PV}, p, n}$ are calculated with the operators $\hat{h}_{\mathrm{SM}, p,n}$ \eqref{eq:h_SI_pn} instead of the full operator $\hat{h}_{\mathrm{SM}}$ \eqref{eq:h_SI_01QED}, see Eq.~\eqref{eq:E_PV}:
\begin{gather}
	\varepsilon^\mathrm{(SM)}_{\mathrm{PV}, p, n}=\sum_k\left(  \frac{\langle 2s|\hat{D}_z|k\rangle \langle k|\hat{h}_{\mathrm{SM}, p,n}|1s\rangle}{E_{1s}-E_k}\right. + \nonumber \\ \left. \frac{\langle 2s|\hat{h}_{\mathrm{SM}, p,n}|k\rangle \langle k|\hat{D}_z|1s\rangle}{E_{2s}-E_k} \right).\label{eq:epsilon_SM}
\end{gather} 
Here the sum runs over the intermediate states $|k\rangle = |np_{1/2}\rangle$.

Furthermore, we rewrite the NP counterparts of the SM matrix elements and amplitudes in a similar way, distinguishing proton and neutron contributions, see Eq.~\eqref{eq:V_integral}:
\begin{align}
		\mathcal{M}^\mathrm{(NP)} &= \frac{\hbar^3}{m_{Z'}^2 c}\left(g_{ep}Z \mathrm{m}^\mathrm{(NP)}_{p} + g_{en}N \mathrm{m}^\mathrm{(NP)}_{n}\right) ,\label{eq:M_NP_split}\\
	\mathcal{E}^\mathrm{(NP)}_\mathrm{PV} &=  \frac{\hbar^3}{m_{Z'}^2 c}\left(g_{ep}Z \varepsilon^\mathrm{(NP)}_{\mathrm{PV}, p} + g_{en}N \varepsilon^\mathrm{(NP)}_{\mathrm{PV}, n}\right) .\label{eq:E_NP_split}
\end{align}
Here, $g_{ep}$ and $g_{en}$ designate the unknown coupling strengths of a hypothetical $Z'$ boson to protons and neutrons, respectively. The PV matrix elements 
\begin{equation}
	\mathrm{m}^\mathrm{(NP)}_{p,n}=\langle 2s|\hat{h}_{{Z'}\, p,n}|2p_{1/2} \rangle ,\label{eq:me_NP}
\end{equation}
are calculated with the NP partial operators
\begin{equation}
	\hat{h}_{{Z'}\, p,n} = \frac{1}{4\pi}\gamma^5 \left(\frac{m_{Z'}c}{\hbar}\right)^2 \int\frac{e^{-m_{Z'}c\left|\vec{r}-\vec{R}\right|/\hbar}}{| \vec{r}-\vec{R}|}\rho_{p,n}(\vec{R}) d\vec{R}\ ,\label{eq:V_pn}
\end{equation}
which correspond to Eq.~\eqref{eq:V_integral} divided by the square reduced Compton wavelength of the $Z'$ boson, $\lambdabar_{c, Z'}^2=\hbar^2 / (m_{Z'}c)^2$. Likewise, the partial amplitudes are
\begin{gather}
	\varepsilon^\mathrm{(NP)}_{\mathrm{PV}, p, n}=\sum_k\left(  \frac{\langle 2s|\hat{D}_z|k\rangle \langle k|\hat{h}_{{Z'}\, p,n}|1s\rangle}{E_{1s}-E_k}\right. + \nonumber \\ \left. \frac{\langle 2s|\hat{h}_{{Z'}\, p,n}|k\rangle \langle k|\hat{D}_z|1s\rangle}{E_{2s}-E_k} \right).\label{eq:epsilon_NP}
\end{gather}

Up to now, we have discussed how the hydrogenic matrix elements and amplitudes can be split into proton and neutron parts. Similar analysis can be performed for Li-like ions, leading to formulas for the SM and NP total amplitudes equivalent to Eqs.~\eqref{eq:E_SM_split} and \eqref{eq:E_NP_split}. For the Li-like case, the partial amplitudes would read
\begin{gather}
	\varepsilon^\mathrm{(SM)}_{\mathrm{PV}, p, n}=\sum_k\left(  \frac{\langle 1s^23s|\hat{D}_z|k\rangle \langle k|\hat{h}_{\mathrm{SM}, p,n}|1s^22s\rangle}{E_{1s^22s}-E_k}\right. + \nonumber \\ \left. \frac{\langle 1s^23s|\hat{h}_{\mathrm{SM}, p,n}|k\rangle \langle k|\hat{D}_z|1s^22s\rangle}{E_{1s^23s}-E_k} \right),\label{eq:epsilon_SM_Li}\\
	\varepsilon^\mathrm{(NP)}_{\mathrm{PV}, p, n}=\sum_k\left(  \frac{\langle 1s^23s|\hat{D}_z|k\rangle \langle k|\hat{h}_{{Z'}\, p,n}|1s^22s\rangle}{E_{1s^22s}-E_k}\right. + \nonumber \\ \left. \frac{\langle 1s^23s|\hat{h}_{{Z'}\, p,n}|k\rangle \langle k|\hat{D}_z|1s^22s\rangle}{E_{1s^23s}-E_k} \right),\label{eq:epsilon_NP_Li}
\end{gather}
where both sums run over $|k\rangle = |1s^2 np_{1/2}\rangle$ intermediate states.

The results of our calculations are summarized in Tabs.~\ref{tab:H_SM_1},~\ref{tab:Li_SM} for the standard model APV, and in Tabs.~\ref{tab:H_dark_Ca}--\ref{tab:Li_dark} for the hypothetical new $Z'$ boson effects. In both cases, we present the matrix elements and PV amplitudes evaluated with respective partial operators \eqref{eq:h_SI_pn} or \eqref{eq:V_pn}, and include their corresponding uncertainties. 
For the standard model APV results, we also show the total values of $\mathcal{M}^\mathrm{(SM)}$ and $\mathcal{E}^\mathrm{(SM)}_\mathrm{PV}$, as in Eqs.~\eqref{eq:M_SM_split} and \eqref{eq:E_SM_split}. While the uncertainty of matrix elements arises, as discussed above, from the incomplete knowledge of proton and neutron distributions, the errors in the APV amplitudes are primarily due to the uncertainties in the energy difference denominators.

\begin{table}[tb]
\caption{Standard-model spin-independent APV in the $1s\rightarrow 2s$ transition of H-like $^{40,48}$Ca and $^{208}$Pb. Proton ($p$) and neutron ($n$) partial matrix elements $\mathrm{m}^\mathrm{(SM)}_{p,n}$~\eqref{eq:me_SM} and amplitudes $\varepsilon^\mathrm{(SM)}_{\mathrm{PV}, p,n}$~\eqref{eq:epsilon_SM} are calculated with the Hamiltonians $\hat{h}_{\mathrm{SM}\, p,n}$~\eqref{eq:h_SI_pn}. The values of the total leading matrix element $\mathcal{M}^\mathrm{(SM)}$ \eqref{eq:M_SM_split} and total amplitude $\mathcal{E}^\mathrm{(SM)}_\mathrm{PV}$ \eqref{eq:E_SM_split} are derived from $\hat{h}_\mathrm{SM}$~\eqref{eq:h_SI_01QED}. The units of matrix elements and PV amplitudes are $10^{-15} i\alpha^2mc^2$ and $10^{-15} i e a_B$, respectively. \label{tab:H_SM_1}}

\begin{ruledtabular}
\begin{tabular}{l|l|l|l}
       &  $^{40}$Ca	        &  $^{48}$Ca 	        &	     $^{208}$Pb 	    \\
\hline
$\mathrm{m}^\mathrm{(SM)}_{p}$	   &  $-372.14(2)$	    &    $-372.122(3)$	  &    $-826378(24)$   \\
$\mathrm{m}^\mathrm{(SM)}_{n}$	   &  $-372.15(3)$	    &    $-371.98(4)$	  &    $-819257(1792)$  \\ 
$\varepsilon^\mathrm{(SM)}_{\mathrm{PV},p}$	 &  $-950.12(6)$   		&    $-950.08(4)$  	&    $-2667(3)$     \\
$\varepsilon^\mathrm{(SM)}_{\mathrm{PV},n}$	 &  $-950.15(6)$   		&    $-949.7(1)$	  	&     $-2644(7)$  \\
\hline
$\mathcal{M}^\mathrm{(SM)}$	  &   $3.4164(3) \times 10^3$ 	&   $4.886(1)\times 10^3$	&    $4.86(1)\times 10^7$  \\
$\mathcal{E}^\mathrm{(SM)}_{\mathrm{PV}}$	  	&   $8.722(1)\times 10^3$	   &  $1.2475(2)\times 10^4$	&     $1.570(4) \times 10^5$ 

\end{tabular}
\end{ruledtabular}
\end{table}

\begin{table}[tb]
\caption{Standard-model spin-independent APV in the $1s^2 2s\rightarrow 1s^2 3s$ transition of Li-like $^{40,48}$Ca and $^{208}$Pb. Proton ($p$) and neutron ($n$) amplitudes $\varepsilon^\mathrm{(SM)}_{\mathrm{PV},p,n}$~\eqref{eq:epsilon_SM_Li} are calculated using the hamiltonians $\hat{h}_{\mathrm{SM}\, p,n}$~\eqref{eq:h_SI_pn} and the total $\mathcal{E}_\mathrm{PV}$ is determined using $\hat{h}_\mathrm{SM}$ \eqref{eq:h_SI_01QED}. The units are $10^{-15} i e a_B$. \label{tab:Li_SM}}

\begin{ruledtabular}
\begin{tabular}{l|l|l|l}
                     &  $^{40}$Ca	    &  $^{48}$Ca 	    &	 $^{208}$Pb 	    \\
\hline
$\varepsilon^\mathrm{(SM)}_{\mathrm{PV},p}$	 &  $9.129(4)$   &    $9.128(4)$  &    $946(2)$     \\
$\varepsilon^\mathrm{(SM)}_{\mathrm{PV},n}$  &  $9.129(4)$   &    $9.125(4)$	&    $938(3)$  \\
\hline
$\mathcal{E}^\mathrm{(SM)}_{\mathrm{PV}}$ & $-83.80(4)$ & $-119.9(1)$ & $-55707(185)$
\end{tabular}
\end{ruledtabular}
\end{table}


\begin{table*}[tb]
\caption{Spin-independent effects of a hypothetical $Z'$ boson of mass $m_{Z'}$ in the $1s\rightarrow 2s$ transition of H-like $^{40,48}$Ca. Proton ($p$) and neutron ($n$) partial matrix elements $\mathrm{m}^\mathrm{(NP)}_{p,n}$ \eqref{eq:me_NP} and amplitudes $\varepsilon^\mathrm{(NP)}_{\mathrm{PV}, p,n}$ \eqref{eq:epsilon_NP} are calculated with the Hamiltonians $\hat{h}_{{Z'}\, p,n}$~\eqref{eq:V_pn}; see also Eqs.~\eqref{eq:M_NP_split} and \eqref{eq:E_NP_split}. The units of the presented partial matrix elements and PV amplitudes are $i\alpha^2 /a_B^3$ and $iem\alpha^2/\hbar^2$, respectively. \label{tab:H_dark_Ca}}
\begin{ruledtabular}
\begin{tabular}{l|l|l|l|l|l|l}
 & \multicolumn{2}{c|}{$^{40}$Ca ($p$)} & \multicolumn{2}{c|}{$^{40}$Ca ($n$)} & \multicolumn{2}{c}{$^{48}$Ca ($p$)} \\
\hline                                                         
 $m_{Z'}$, eV &  $\mathrm{m}^\mathrm{(NP)}_{p}$ 	    &  $\varepsilon^\mathrm{(NP)}_{\mathrm{PV}, p}$	                     &  $\mathrm{m}^\mathrm{(NP)}_{n}$ 	          &  $\varepsilon^\mathrm{(NP)}_{\mathrm{PV}, n}$	                      &  $\mathrm{m}^\mathrm{(NP)}_{p}$ 	               &    $\varepsilon^\mathrm{(NP)}_{\mathrm{PV}, p}$           \\
 \hline                  
$10^{1}$ & $-1.2329\times 10^{-7}$  & $-3.1466(1)\times 10^{-7}$ &  $-1.2329\times 10^{-7}$  & $-3.1466(1)\times 10^{-7}$  &  $-1.2329\times 10^{-7}$ & $-3.1466(1)\times 10^{-7}$        \\                
$10^{2}$ & $-1.2329\times 10^{-5}$  & $-3.1466(1)\times 10^{-5}$ &  $-1.2329\times 10^{-5}$  & $-3.1466(1)\times 10^{-5}$  &  $-1.2329\times 10^{-5}$ & $-3.1466(1)\times 10^{-5}$       \\                
$10^{3}$ & $-1.2323\times 10^{-3}$  & $-3.1451(1)\times 10^{-3}$ &  $-1.2323\times 10^{-3}$  & $-3.1451(1)\times 10^{-3}$  &  $-1.2323\times 10^{-3}$ & $-3.1451(1)\times 10^{-3}$       \\                 
$10^{4}$ & $-0.11862 $              & $-0.30280(1)$              &  $-0.11862 $              & $-0.30280(1)$               &  $-0.11862 $             & $-0.30280(1)$                    \\                 
$10^{5}$ & $-4.9000	$               & $-12.5099(6)	$            &  $-4.9000	$              & $-12.5099(6)	$              &  $-4.9000	$             & $-12.5100(5)	$                   \\                 
$10^{6}$ & $-17.950	$               & $-45.829(2)	$              &  $-17.950	$              & $-45.829(2)	$              &  $-17.950	$             & $-45.830(2)	$                    \\                 
$10^{7}$ & $-22.339	$               & $-57.033(3)	$              &  $-22.339	$              & $-57.034(3)	$              &  $-22.339	$             & $-57.034(2)	$                    \\                 
$10^{8}$ & $-23.534(1)$             & $-60.085(3)	$              &  $-23.534(1)	$            & $-60.086(3)	$              &  $-23.5335(1)	$         & $-60.084(3)	$                    \\                 
$10^{9}$ & $-23.677(1)$	            & $-60.451(4)	$              &  $-23.678(2)$	           & $-60.453(4)	$              &  $-23.6763(2)$	          & $-60.449(3)	$                                               
\end{tabular}
\end{ruledtabular}
\end{table*}

\begin{table*}[tb]
\caption{Spin-independent effects of a hypothetical $Z'$ boson of mass $m_{Z'}$ in the $1s\rightarrow 2s$ transition of H-like $^{48}$Ca and $^{208}$Pb. Proton ($p$) and neutron ($n$) partial matrix elements $\mathrm{m}^\mathrm{(NP)}_{p,n}$ \eqref{eq:me_NP} and amplitudes $\varepsilon^\mathrm{(NP)}_{\mathrm{PV}, p,n}$ \eqref{eq:epsilon_NP} are calculated with the Hamiltonians $\hat{h}_{{Z'}\, p,n}$~\eqref{eq:V_pn}; see also Eqs.~\eqref{eq:M_NP_split} and \eqref{eq:E_NP_split}. The units of the presented partial matrix elements and PV amplitudes are $i\alpha^2 /a_B^3$ and $iem\alpha^2/\hbar^2$, respectively.  \label{tab:H_dark_Pb}}
\begin{ruledtabular}
\begin{tabular}{l|l|l|l|l|l|l}
 & \multicolumn{2}{c|}{$^{48}$Ca ($n$)} & \multicolumn{2}{c|}{$^{208}$Pb ($p$)} & \multicolumn{2}{c}{$^{208}$Pb ($n$)} \\
\hline                                                         
 $m_{Z'}$, eV &  $\mathrm{m}^\mathrm{(NP)}_{n}$ 	    &  $\varepsilon^\mathrm{(NP)}_{\mathrm{PV}, n}$	                     &  $\mathrm{m}^\mathrm{(NP)}_{p}$ 	          &  $\varepsilon^\mathrm{(NP)}_{\mathrm{PV}, p}$	                      &  $\mathrm{m}^\mathrm{(NP)}_{n}$ 	               &    $\varepsilon^\mathrm{(NP)}_{\mathrm{PV}, n}$           \\
 \hline  
\hline
$10^{1}$      & $-1.2329\times 10^{-7}$  & $-3.1466(1)\times 10^{-7}$  &  $-3.1737\times 10^{-6}$   & $-1.022(1)\times 10^{-8}$  &  $-3.1736\times 10^{-6}$   & $-1.022(1)\times 10^{-8}$     \\                
$10^{2}$      & $-1.2329\times 10^{-5}$  & $-3.1466(1)\times 10^{-5}$  &  $-3.1737\times 10^{-4}$   & $-1.022(1)\times 10^{-6}$  &  $-3.1736\times 10^{-4}$   & $-1.022(1)\times 10^{-6}$    \\                
$10^{3}$      & $-1.2323\times 10^{-3}$  & $-3.1451(1)\times 10^{-3}$  &  $-3.1736\times 10^{-2}$   & $-1.022(1)\times 10^{-4}$  &  $-3.1735\times 10^{-2}$   & $-1.022(1)\times 10^{-4}$    \\                 
$10^{4}$      & $-0.11862 $              & $-0.30280(1)$               &  $-3.1675 $                & $-0.01020(1) $             &  $-3.1675 $                & $-0.01020(1) $               \\                 
$10^{5}$      & $-4.9000	$              & $-12.5099(5)	$              &  $-283.73	$               & $-0.915(1)	$              &  $-283.73	$               & $-0.915(1)	$                 \\                 
$10^{6}$      & $-17.950	$              & $-45.830(2)	$              &  $-7268.6	$               & $-23.45(3)	$              &  $-7267.9(2)	$             & $-23.45(3)	$                 \\                 
$10^{7}$      & $-22.3382(1)$            & $-57.032(2)	$              &  $-29417.4(3)	$           & $-94.9(1)	$                &  $-29384(8)	$             & $-94.8(1)	$                  \\                 
$10^{8}$      & $-23.528(2)$             & $-60.070(5)	$              &  $-50330(1)	$             & $-162.4(2)	$              &  $-49985(87)	$             & $-161.3(3)	$                 \\                 
$10^{9}$      & $-23.667(3)$	           & $-60.426(7)	$              &  $-52555(2)$	              & $-169.6(2)	$              &  $-52103(114)$	            & $-168.1(4)	$                         
\end{tabular}
\end{ruledtabular}
\end{table*}

\begin{table*}[tb]
\caption{Spin-independent effects of a hypothetical new $Z'$ boson of mass $m_{Z'}$ in the $1s^22s\rightarrow 1s^23s$ transition of Li-like $^{40,48}$Ca and $^{208}$Pb. Proton ($p$) and neutron ($n$) partial amplitudes $\varepsilon^\mathrm{(NP)}_{\mathrm{PV}, p,n}$ \eqref{eq:epsilon_NP_Li} are calculated with the Hamiltonians $\hat{h}_{{Z'}\, p,n}$~\eqref{eq:V_pn}; see also Eqs.~\eqref{eq:M_NP_split} and \eqref{eq:E_NP_split}.
The units of $\varepsilon^\mathrm{(NP)}_{\mathrm{PV}, p,n}$ are $iem\alpha^2/\hbar^2$.\label{tab:Li_dark}}
\begin{ruledtabular}
\begin{tabular}{l|l|l|l|l|l|l}
 & $^{40}$Ca ($p$) & $^{40}$Ca ($n$) & $^{48}$Ca ($p$) & $^{48}$Ca ($n$) & $^{208}$Pb ($p$) & $^{208}$Pb ($n$) \\
\hline
$m_{Z'}$, eV  & $\varepsilon^\mathrm{(NP)}_{\mathrm{PV},p}$ & $\varepsilon^\mathrm{(NP)}_{\mathrm{PV},n}$ & $\varepsilon^\mathrm{(NP)}_{\mathrm{PV},p}$ & $\varepsilon^\mathrm{(NP)}_{\mathrm{PV},n}$ & $\varepsilon^\mathrm{(NP)}_{\mathrm{PV},p}$ & $\varepsilon^\mathrm{(NP)}_{\mathrm{PV},n}$\\
\hline
                   
$10^{1}$ & $3.797(1)\times 10^{-9}$     & $3.797(1)\times 10^{-9}$ 	& $3.797(1)\times 10^{-9}$   &  $3.797(1)\times 10^{-9}$	& $4.055(9)\times 10^{-9}$ 	   & $4.054(9)\times 10^{-9}$  \\                
$10^{2}$ & $3.795(1)\times 10^{-7}$     & $3.795(1)\times 10^{-7}$ 	& $3.795(1)\times 10^{-7}$ 	 &  $3.795(1)\times 10^{-7}$	& $4.055(9)\times 10^{-7}$ 	   & $4.054(9)\times 10^{-7}$  \\                
$10^{3}$ & $3.771(1)\times 10^{-5}$     & $3.771(1)\times 10^{-5}$ 	& $3.771(1)\times 10^{-5}$ 	 &  $3.771(1)\times 10^{-5}$	& $4.054(9)\times 10^{-5}$ 	   & $4.054(9)\times 10^{-5}$  \\                 
$10^{4}$ & $3.340(1)\times 10^{-3}$     & $3.340(1)\times 10^{-3}$	& $3.340(1)\times 10^{-3}$ 	 &  $3.340(1)\times 10^{-3}$	& $4.029(9)\times 10^{-3}$ 	   & $4.029(9)\times 10^{-3}$  \\               
$10^{5}$ & $0.11816(5)$ 					& $0.11816(5)$ 				& $0.11816(5)$ 				 &  $0.11816(5)$ 			& $0.3244(7)$ 				   & $0.3244(7)$ 				\\                 
$10^{6}$ & $0.4400(2)$ 					& $0.4400(2)$ 				& $0.4400(2)$ 				 &  $0.4400(2)$ 				& $8.12(2)$ 				       & $8.12(2)$ 				    \\                
$10^{7}$ & $0.5480(2)$ 					& $0.5480(2)$ 				& $0.5480(2)$ 				 &  $0.5480(2)$ 				& $33.63(8)$	 				   & $33.59(8)$	 				\\           
$10^{8}$ & $0.5773(2)$ 					& $0.5773(2)$ 				& $0.5773(2)$ 				 &  $0.5772(2)$ 				& $57.6(1)$ 				       & $57.2(2)$ 				    \\              
$10^{9}$ & $0.5808(2)$ 					& $0.5808(2)$ 				& $0.5808(2)$ 				 &  $0.5806(2)$ 				& $60.2(1)$ 				       & $59.7(2)$ 				          			
\end{tabular}
\end{ruledtabular}
\end{table*}

From the comparison of the matrix elements $\mathrm{m}_{p}^\mathrm{(SM)}$ and $\mathrm{m}_{n}^\mathrm{(SM)}$ as well as the amplitudes $\varepsilon_{p}^\mathrm{(SM)}$ and $\varepsilon_{n}^\mathrm{(SM)}$ in Tabs.~\ref{tab:H_SM_1},~\ref{tab:Li_SM}, it is evident that the neutron skin contribution to the standard model APV is much stronger in $^{208}$Pb than in $^{40,48}$Ca. For example, in $^{48}$Ca, the difference between $\varepsilon_{p}^\mathrm{(SM)}$ and $\varepsilon_{n}^\mathrm{(SM)}$ is 0.03\%, while in $^{208}$Pb it is 0.8\%. This is a consequence not only of the thicker neutron skin in $^{208}$Pb \eqref{eq:Pb_nsk}, but also of the higher nuclear charge $Z$ \cite{budkerAtomicPhysicsStudies2020}. The relatively large effect in $^{208}$Pb would potentially allow for a neutron skin detection through APV in highly charged $^{208}$Pb ions, a measurement complementary to the PREX-II experiment \cite{prexcollaborationAccurateDeterminationNeutron2021}. 

On the other hand, the APV effect of a hypothetical light $Z'$ boson is not particularly sensitive to the neutron skin. Again, this can be seen from the matrix elements $\mathrm{m}^\mathrm{(NP)}_{p,n}$ and amplitudes $\varepsilon^\mathrm{(NP)}_{p,n}$ for different masses $m_{Z'}$ of a hypothetical new physics boson, which are presented in Tabs.\ref{tab:H_dark_Ca}--\ref{tab:Li_dark}. As seen from these tables, neutron skin effects begin to matter only for masses exceeding $m_{Z'}= 0.1$~GeV. This fact can be readily explained by the consideration that lighter bosons have larger Compton wavelength and therefore cannot `sense' the fine distinction between proton and neutron distributions. And even for a heavy boson case the difference between $\varepsilon^\mathrm{(NP)}_{n}$ and $\varepsilon^\mathrm{(NP)}_{p}$ is below 0.9\% for $^{208}$Pb and even smaller for $^{40,48}$Ca, which is similar to the SM APV effect.

The weak sensitivity of PV new physics amplitudes to the neutron skin effects in the case of $^{40,48}$Ca ions constitutes an advantage for the search of new physics. When the neutron skin is not `felt' by the lighter $Z'$ bosons and has only a minuscule effect upon the standard model APV, one can disregard the nuclear distribution error, which is otherwise a major source of uncertainty in isotopic differences \eqref{eq:Ca_M_difference} and ratios \eqref{eq:Ca_M_ratio}. Hence, for the search of new physics, one may assume that proton and neutron densities in Ca are equal, i.e., the neutron skin $\Delta r_\mathrm{rms}=0$. Moreover, since the root-mean-square charge radii $r_p$ in $^{40}$Ca and $^{48}$Ca are similar (see Tab.~\ref{tab:nuc_param}), it is safe to assume that $r_p(^{40}\mathrm{Ca})=r_n(^{40}\mathrm{Ca})=r_p(^{48}\mathrm{Ca})=r_n(^{48}\mathrm{Ca})$. At the same time, the neutron difference $\Delta N=8$ between the $^{40,48}$Ca nuclei is significant, which leads to the total value of the SM APV effect being 1.5 times larger in $^{48}$Ca than in $^{40}$Ca. In experiment, these properties of the $^{40,48}$Ca isotope pair allow for a clear separation between the hypothetical $Z'$-boson coupling strengths to protons and neutrons. Furthermore, the unique combination of almost equal proton and neutron radii with a relatively large neutron number difference translates into an advantage for the search of new physics in neutral or multiply-charged $^{40,48}$Ca as well. However, given the $\sim Z^3$ scaling of the APV effects in neutral atoms and multiply charged ions \cite{bouchiatParityViolationInduced1974}, the parity-violation measurements in Ca$^{+}$ are rarely considered, in comparison to those in heavier atoms.

\section{Conclusion and Outlook}\label{sec:conclusion}
In summary, we present a theoretical investigation of atomic parity violation amplitudes in H- and Li-like ions. Special attention is paid to the effects that arise from neutron skin and hypothetical new physics particles. To explore these effects, detailed relativistic calculations for H- and Li-like $^{40,48}$Ca and $^{208}$Pb ions have been performed. Based on their results, we argue that the isotope pair $^{40}$Ca and $^{48}$Ca provides a unique testing ground for the search for new physics. That is because the PV amplitude in this isotope pair is barely sensitive to the neutron skin effects, and the nuclear charge radii $r_p(^{40}\mathrm{Ca})$ and $r_p(^{48}\mathrm{Ca})$ are almost equal. On the other hand, a prominent neutron skin effect is predicted for $^{208}$Pb, which makes highly charged ions of this isotope favorable for neutron skin detection by means of APV, in a way complementary to the PREX-II \cite{prexcollaborationAccurateDeterminationNeutron2021} experiment.

In the present work, we have focused on even-even spinless isotopes and have therefore not considered nuclear-spin-dependent (NSD) APV effects. Even though they are typically of the order of 1\% of the absolute size of the nuclear-spin-independent APV, NSD effects are expected to be measurable in HCIs as well, for instance in the hyperfine transitions of He-like ions with closely lying 2$^3P_1$ and 2$^1S_0$ states, such as $^{77}_{34}$Se \cite{ferroHyperfineTransitionsHelike2011,budkerAtomicPhysicsStudies2020}. The NSD APV measurements in HCIs are intriguing because they could provide a test for potential new spin-dependent parity-violating interactions, as well as insights into the quadrupole deformation of nuclear neutron distributions \cite{flambaumEffectNuclearQuadrupole2017}. The analysis and prospects of such measurements will be addressed in our future publications.

\section{Acknowledgements}
The present work was funded by Deutsche Forschungsgemeinschaft (DFG, German Research Foundation) through the projects 544815538 and Germany’s Excellence Strategy--EXC--2123
QuantumFrontiers--390837967. We are grateful to Jan Richter and Jonas Sommerfeld for insightful discussions.


\begin{thebibliography}{32}%
\makeatletter
\providecommand \@ifxundefined [1]{%
 \@ifx{#1\undefined}
}%
\providecommand \@ifnum [1]{%
 \ifnum #1\expandafter \@firstoftwo
 \else \expandafter \@secondoftwo
 \fi
}%
\providecommand \@ifx [1]{%
 \ifx #1\expandafter \@firstoftwo
 \else \expandafter \@secondoftwo
 \fi
}%
\providecommand \natexlab [1]{#1}%
\providecommand \enquote  [1]{``#1''}%
\providecommand \bibnamefont  [1]{#1}%
\providecommand \bibfnamefont [1]{#1}%
\providecommand \citenamefont [1]{#1}%
\providecommand \href@noop [0]{\@secondoftwo}%
\providecommand \href [0]{\begingroup \@sanitize@url \@href}%
\providecommand \@href[1]{\@@startlink{#1}\@@href}%
\providecommand \@@href[1]{\endgroup#1\@@endlink}%
\providecommand \@sanitize@url [0]{\catcode `\\12\catcode `\$12\catcode `\&12\catcode `\#12\catcode `\^12\catcode `\_12\catcode `\%12\relax}%
\providecommand \@@startlink[1]{}%
\providecommand \@@endlink[0]{}%
\providecommand \url  [0]{\begingroup\@sanitize@url \@url }%
\providecommand \@url [1]{\endgroup\@href {#1}{\urlprefix }}%
\providecommand \urlprefix  [0]{URL }%
\providecommand \Eprint [0]{\href }%
\providecommand \doibase [0]{https://doi.org/}%
\providecommand \selectlanguage [0]{\@gobble}%
\providecommand \bibinfo  [0]{\@secondoftwo}%
\providecommand \bibfield  [0]{\@secondoftwo}%
\providecommand \translation [1]{[#1]}%
\providecommand \BibitemOpen [0]{}%
\providecommand \bibitemStop [0]{}%
\providecommand \bibitemNoStop [0]{.\EOS\space}%
\providecommand \EOS [0]{\spacefactor3000\relax}%
\providecommand \BibitemShut  [1]{\csname bibitem#1\endcsname}%
\let\auto@bib@innerbib\@empty
\bibitem [{\citenamefont {Bouchiat}\ and\ \citenamefont {Bouchiat}(1997)}]{bouchiatParityViolationAtoms1997}%
  \BibitemOpen
  \bibfield  {author} {\bibinfo {author} {\bibfnamefont {M.-A.}\ \bibnamefont {Bouchiat}}\ and\ \bibinfo {author} {\bibfnamefont {C.}~\bibnamefont {Bouchiat}},\ }\bibfield  {title} {\bibinfo {title} {Parity violation in atoms},\ }\href {https://doi.org/10.1088/0034-4885/60/11/004} {\bibfield  {journal} {\bibinfo  {journal} {Reports on Progress in Physics}\ }\textbf {\bibinfo {volume} {60}},\ \bibinfo {pages} {1351} (\bibinfo {year} {1997})}\BibitemShut {NoStop}%
\bibitem [{\citenamefont {Ginges}\ and\ \citenamefont {Flambaum}(2004)}]{gingesViolationsFundamentalSymmetries2004}%
  \BibitemOpen
  \bibfield  {author} {\bibinfo {author} {\bibfnamefont {J.~S.~M.}\ \bibnamefont {Ginges}}\ and\ \bibinfo {author} {\bibfnamefont {V.~V.}\ \bibnamefont {Flambaum}},\ }\bibfield  {title} {\bibinfo {title} {Violations of fundamental symmetries in atoms and tests of unification theories of elementary particles},\ }\href {https://doi.org/10.1016/j.physrep.2004.03.005} {\bibfield  {journal} {\bibinfo  {journal} {Physics Reports}\ }\textbf {\bibinfo {volume} {397}},\ \bibinfo {pages} {63} (\bibinfo {year} {2004})}\BibitemShut {NoStop}%
\bibitem [{\citenamefont {Derevianko}\ and\ \citenamefont {Porsev}(2007)}]{dereviankoTheoreticalOverviewAtomic2007}%
  \BibitemOpen
  \bibfield  {author} {\bibinfo {author} {\bibfnamefont {A.}~\bibnamefont {Derevianko}}\ and\ \bibinfo {author} {\bibfnamefont {S.~G.}\ \bibnamefont {Porsev}},\ }\bibfield  {title} {\bibinfo {title} {Theoretical overview of atomic parity violation},\ }\href {https://doi.org/10.1140/epja/i2006-10427-7} {\bibfield  {journal} {\bibinfo  {journal} {The European Physical Journal A}\ }\textbf {\bibinfo {volume} {32}},\ \bibinfo {pages} {517} (\bibinfo {year} {2007})}\BibitemShut {NoStop}%
\bibitem [{\citenamefont {Roberts}\ \emph {et~al.}(2015)\citenamefont {Roberts}, \citenamefont {Dzuba},\ and\ \citenamefont {Flambaum}}]{robertsParityTimeReversalViolation2015}%
  \BibitemOpen
  \bibfield  {author} {\bibinfo {author} {\bibfnamefont {B.~M.}\ \bibnamefont {Roberts}}, \bibinfo {author} {\bibfnamefont {V.~A.}\ \bibnamefont {Dzuba}},\ and\ \bibinfo {author} {\bibfnamefont {V.~V.}\ \bibnamefont {Flambaum}},\ }\bibfield  {title} {\bibinfo {title} {Parity and {{Time-Reversal Violation}} in {{Atomic Systems}}},\ }\href {https://doi.org/10.1146/annurev-nucl-102014-022331} {\bibfield  {journal} {\bibinfo  {journal} {Annual Review of Nuclear and Particle Science}\ }\textbf {\bibinfo {volume} {65}},\ \bibinfo {pages} {63} (\bibinfo {year} {2015})}\BibitemShut {NoStop}%
\bibitem [{\citenamefont {Navas}\ \emph {et~al.}(2024)\citenamefont {Navas} \emph {et~al.}}]{ParticleDataGroup:2024cfk}%
  \BibitemOpen
  \bibfield  {author} {\bibinfo {author} {\bibfnamefont {S.}~\bibnamefont {Navas}} \emph {et~al.} (\bibinfo {collaboration} {Particle Data Group}),\ }\bibfield  {title} {\bibinfo {title} {Review of particle physics},\ }\href {https://doi.org/10.1103/PhysRevD.110.030001} {\bibfield  {journal} {\bibinfo  {journal} {Physical Review D: Particles and Fields}\ }\textbf {\bibinfo {volume} {110}},\ \bibinfo {pages} {030001} (\bibinfo {year} {2024})}\BibitemShut {NoStop}%
\bibitem [{\citenamefont {Safronova}\ \emph {et~al.}(2018)\citenamefont {Safronova}, \citenamefont {Budker}, \citenamefont {DeMille}, \citenamefont {Kimball}, \citenamefont {Derevianko},\ and\ \citenamefont {Clark}}]{safronovaSearchNewPhysics2018}%
  \BibitemOpen
  \bibfield  {author} {\bibinfo {author} {\bibfnamefont {M.~S.}\ \bibnamefont {Safronova}}, \bibinfo {author} {\bibfnamefont {D.}~\bibnamefont {Budker}}, \bibinfo {author} {\bibfnamefont {D.}~\bibnamefont {DeMille}}, \bibinfo {author} {\bibfnamefont {D.~F.~J.}\ \bibnamefont {Kimball}}, \bibinfo {author} {\bibfnamefont {A.}~\bibnamefont {Derevianko}},\ and\ \bibinfo {author} {\bibfnamefont {C.~W.}\ \bibnamefont {Clark}},\ }\bibfield  {title} {\bibinfo {title} {Search for new physics with atoms and molecules},\ }\href {https://doi.org/10.1103/RevModPhys.90.025008} {\bibfield  {journal} {\bibinfo  {journal} {Reviews of Modern Physics}\ }\textbf {\bibinfo {volume} {90}},\ \bibinfo {pages} {025008} (\bibinfo {year} {2018})}\BibitemShut {NoStop}%
\bibitem [{\citenamefont {Dzuba}\ \emph {et~al.}(2017)\citenamefont {Dzuba}, \citenamefont {Flambaum},\ and\ \citenamefont {Stadnik}}]{dzubaProbingLowMassVector2017}%
  \BibitemOpen
  \bibfield  {author} {\bibinfo {author} {\bibfnamefont {V.~A.}\ \bibnamefont {Dzuba}}, \bibinfo {author} {\bibfnamefont {V.~V.}\ \bibnamefont {Flambaum}},\ and\ \bibinfo {author} {\bibfnamefont {Y.~V.}\ \bibnamefont {Stadnik}},\ }\bibfield  {title} {\bibinfo {title} {Probing {{Low-Mass Vector Bosons}} with {{Parity Nonconservation}} and {{Nuclear Anapole Moment Measurements}} in {{Atoms}} and {{Molecules}}},\ }\href {https://doi.org/10.1103/PhysRevLett.119.223201} {\bibfield  {journal} {\bibinfo  {journal} {Physical Review Letters}\ }\textbf {\bibinfo {volume} {119}},\ \bibinfo {pages} {223201} (\bibinfo {year} {2017})}\BibitemShut {NoStop}%
\bibitem [{\citenamefont {Wood}\ \emph {et~al.}(1997)\citenamefont {Wood}, \citenamefont {Bennett}, \citenamefont {Cho}, \citenamefont {Masterson}, \citenamefont {Roberts}, \citenamefont {Tanner},\ and\ \citenamefont {Wieman}}]{woodMeasurementParityNonconservation1997}%
  \BibitemOpen
  \bibfield  {author} {\bibinfo {author} {\bibfnamefont {C.~S.}\ \bibnamefont {Wood}}, \bibinfo {author} {\bibfnamefont {S.~C.}\ \bibnamefont {Bennett}}, \bibinfo {author} {\bibfnamefont {D.}~\bibnamefont {Cho}}, \bibinfo {author} {\bibfnamefont {B.~P.}\ \bibnamefont {Masterson}}, \bibinfo {author} {\bibfnamefont {J.~L.}\ \bibnamefont {Roberts}}, \bibinfo {author} {\bibfnamefont {C.~E.}\ \bibnamefont {Tanner}},\ and\ \bibinfo {author} {\bibfnamefont {C.~E.}\ \bibnamefont {Wieman}},\ }\bibfield  {title} {\bibinfo {title} {Measurement of {{Parity Nonconservation}} and an {{Anapole Moment}} in {{Cesium}}},\ }\href {https://doi.org/10.1126/science.275.5307.1759} {\bibfield  {journal} {\bibinfo  {journal} {Science}\ }\textbf {\bibinfo {volume} {275}},\ \bibinfo {pages} {1759} (\bibinfo {year} {1997})}\BibitemShut {NoStop}%
\bibitem [{\citenamefont {Antypas}\ \emph {et~al.}(2019)\citenamefont {Antypas}, \citenamefont {Fabricant}, \citenamefont {Stalnaker}, \citenamefont {Tsigutkin}, \citenamefont {Flambaum},\ and\ \citenamefont {Budker}}]{antypasIsotopicVariationParity2019}%
  \BibitemOpen
  \bibfield  {author} {\bibinfo {author} {\bibfnamefont {D.}~\bibnamefont {Antypas}}, \bibinfo {author} {\bibfnamefont {A.}~\bibnamefont {Fabricant}}, \bibinfo {author} {\bibfnamefont {J.~E.}\ \bibnamefont {Stalnaker}}, \bibinfo {author} {\bibfnamefont {K.}~\bibnamefont {Tsigutkin}}, \bibinfo {author} {\bibfnamefont {V.~V.}\ \bibnamefont {Flambaum}},\ and\ \bibinfo {author} {\bibfnamefont {D.}~\bibnamefont {Budker}},\ }\bibfield  {title} {\bibinfo {title} {Isotopic variation of parity violation in atomic ytterbium},\ }\href {https://doi.org/10.1038/s41567-018-0312-8} {\bibfield  {journal} {\bibinfo  {journal} {Nature Physics}\ }\textbf {\bibinfo {volume} {15}},\ \bibinfo {pages} {120} (\bibinfo {year} {2019})}\BibitemShut {NoStop}%
\bibitem [{\citenamefont {Barkov}\ and\ \citenamefont {Zolotorev}(1979)}]{barkovParityViolationAtomic1979}%
  \BibitemOpen
  \bibfield  {author} {\bibinfo {author} {\bibfnamefont {L.~M.}\ \bibnamefont {Barkov}}\ and\ \bibinfo {author} {\bibfnamefont {M.~S.}\ \bibnamefont {Zolotorev}},\ }\bibfield  {title} {\bibinfo {title} {Parity violation in atomic bismuth},\ }\href {https://doi.org/10.1016/0370-2693(79)90604-X} {\bibfield  {journal} {\bibinfo  {journal} {Physics Letters B}\ }\textbf {\bibinfo {volume} {85}},\ \bibinfo {pages} {308} (\bibinfo {year} {1979})}\BibitemShut {NoStop}%
\bibitem [{\citenamefont {Krasny}(2015)}]{krasnyGammaFactoryProposal2015}%
  \BibitemOpen
  \bibfield  {author} {\bibinfo {author} {\bibfnamefont {M.~W.}\ \bibnamefont {Krasny}},\ }\href {https://doi.org/10.48550/arXiv.1511.07794} {\bibinfo {title} {The {{Gamma Factory}} proposal for {{CERN}}}} (\bibinfo {year} {2015}),\ \Eprint {https://arxiv.org/abs/1511.07794} {arXiv:1511.07794 [hep-ex]} \BibitemShut {NoStop}%
\bibitem [{\citenamefont {Richter}\ \emph {et~al.}(2022)\citenamefont {Richter}, \citenamefont {Maiorova}, \citenamefont {Viatkina}, \citenamefont {Budker},\ and\ \citenamefont {Surzhykov}}]{richterParityViolationStudiesPartially2022}%
  \BibitemOpen
  \bibfield  {author} {\bibinfo {author} {\bibfnamefont {J.}~\bibnamefont {Richter}}, \bibinfo {author} {\bibfnamefont {A.~V.}\ \bibnamefont {Maiorova}}, \bibinfo {author} {\bibfnamefont {A.~V.}\ \bibnamefont {Viatkina}}, \bibinfo {author} {\bibfnamefont {D.}~\bibnamefont {Budker}},\ and\ \bibinfo {author} {\bibfnamefont {A.}~\bibnamefont {Surzhykov}},\ }\bibfield  {title} {\bibinfo {title} {Parity-{{Violation Studies}} with {{Partially Stripped Ions}}},\ }\href {https://doi.org/10.1002/andp.202100561} {\bibfield  {journal} {\bibinfo  {journal} {Annalen der Physik}\ }\textbf {\bibinfo {volume} {534}},\ \bibinfo {pages} {2100561} (\bibinfo {year} {2022})}\BibitemShut {NoStop}%
\bibitem [{\citenamefont {Budker}\ \emph {et~al.}(2020)\citenamefont {Budker}, \citenamefont {{Crespo L{\'o}pez-Urrutia}}, \citenamefont {Derevianko}, \citenamefont {Flambaum}, \citenamefont {Krasny}, \citenamefont {Petrenko}, \citenamefont {Pustelny}, \citenamefont {Surzhykov}, \citenamefont {Yerokhin},\ and\ \citenamefont {Zolotorev}}]{budkerAtomicPhysicsStudies2020}%
  \BibitemOpen
  \bibfield  {author} {\bibinfo {author} {\bibfnamefont {D.}~\bibnamefont {Budker}}, \bibinfo {author} {\bibfnamefont {J.~R.}\ \bibnamefont {{Crespo L{\'o}pez-Urrutia}}}, \bibinfo {author} {\bibfnamefont {A.}~\bibnamefont {Derevianko}}, \bibinfo {author} {\bibfnamefont {V.~V.}\ \bibnamefont {Flambaum}}, \bibinfo {author} {\bibfnamefont {M.~W.}\ \bibnamefont {Krasny}}, \bibinfo {author} {\bibfnamefont {A.}~\bibnamefont {Petrenko}}, \bibinfo {author} {\bibfnamefont {S.}~\bibnamefont {Pustelny}}, \bibinfo {author} {\bibfnamefont {A.}~\bibnamefont {Surzhykov}}, \bibinfo {author} {\bibfnamefont {V.~A.}\ \bibnamefont {Yerokhin}},\ and\ \bibinfo {author} {\bibfnamefont {M.}~\bibnamefont {Zolotorev}},\ }\bibfield  {title} {\bibinfo {title} {Atomic {{Physics Studies}} at the {{Gamma Factory}} at {{CERN}}},\ }\href {https://doi.org/10.1002/andp.202000204} {\bibfield  {journal} {\bibinfo  {journal} {Annalen der Physik}\ }\textbf {\bibinfo {volume} {532}},\ \bibinfo {pages} {2000204} (\bibinfo {year} {2020})}\BibitemShut {NoStop}%
\bibitem [{\citenamefont {Thiel}\ \emph {et~al.}(2019)\citenamefont {Thiel}, \citenamefont {Sfienti}, \citenamefont {Piekarewicz}, \citenamefont {Horowitz},\ and\ \citenamefont {Vanderhaeghen}}]{thielNeutronSkinsAtomic2019}%
  \BibitemOpen
  \bibfield  {author} {\bibinfo {author} {\bibfnamefont {M.}~\bibnamefont {Thiel}}, \bibinfo {author} {\bibfnamefont {C.}~\bibnamefont {Sfienti}}, \bibinfo {author} {\bibfnamefont {J.}~\bibnamefont {Piekarewicz}}, \bibinfo {author} {\bibfnamefont {C.~J.}\ \bibnamefont {Horowitz}},\ and\ \bibinfo {author} {\bibfnamefont {M.}~\bibnamefont {Vanderhaeghen}},\ }\bibfield  {title} {\bibinfo {title} {Neutron skins of atomic nuclei: Per aspera ad astra},\ }\href {https://doi.org/10.1088/1361-6471/ab2c6d} {\bibfield  {journal} {\bibinfo  {journal} {Journal of Physics G: Nuclear and Particle Physics}\ }\textbf {\bibinfo {volume} {46}},\ \bibinfo {pages} {093003} (\bibinfo {year} {2019})}\BibitemShut {NoStop}%
\bibitem [{\citenamefont {Adhikari}\ \emph {et~al.}(2022)\citenamefont {Adhikari} \emph {et~al.}}]{crexcollaborationPrecisionDeterminationNeutral2022}%
  \BibitemOpen
  \bibfield  {author} {\bibinfo {author} {\bibfnamefont {D.}~\bibnamefont {Adhikari}} \emph {et~al.} (\bibinfo {collaboration} {CREX Collaboration}),\ }\bibfield  {title} {\bibinfo {title} {Precision {{Determination}} of the {{Neutral Weak Form Factor}} of $^{48}${Ca}},\ }\href {https://doi.org/10.1103/PhysRevLett.129.042501} {\bibfield  {journal} {\bibinfo  {journal} {Physical Review Letters}\ }\textbf {\bibinfo {volume} {129}},\ \bibinfo {pages} {042501} (\bibinfo {year} {2022})}\BibitemShut {NoStop}%
\bibitem [{\citenamefont {Adhikari}\ \emph {et~al.}(2021)\citenamefont {Adhikari} \emph {et~al.}}]{prexcollaborationAccurateDeterminationNeutron2021}%
  \BibitemOpen
  \bibfield  {author} {\bibinfo {author} {\bibfnamefont {D.}~\bibnamefont {Adhikari}} \emph {et~al.} (\bibinfo {collaboration} {{PREX Collaboration}}),\ }\bibfield  {title} {\bibinfo {title} {Accurate {{Determination}} of the {{Neutron Skin Thickness}} of $^{208}${Pb} through {{Parity-Violation}} in {{Electron Scattering}}},\ }\href {https://doi.org/10.1103/PhysRevLett.126.172502} {\bibfield  {journal} {\bibinfo  {journal} {Physical Review Letters}\ }\textbf {\bibinfo {volume} {126}},\ \bibinfo {pages} {172502} (\bibinfo {year} {2021})}\BibitemShut {NoStop}%
\bibitem [{\citenamefont {Erler}\ and\ \citenamefont {Su}(2013)}]{erlerWeakNeutralCurrent2013}%
  \BibitemOpen
  \bibfield  {author} {\bibinfo {author} {\bibfnamefont {J.}~\bibnamefont {Erler}}\ and\ \bibinfo {author} {\bibfnamefont {S.}~\bibnamefont {Su}},\ }\bibfield  {title} {\bibinfo {title} {The weak neutral current},\ }\href {https://doi.org/10.1016/j.ppnp.2013.03.004} {\bibfield  {journal} {\bibinfo  {journal} {Progress in Particle and Nuclear Physics}\ }\bibinfo {series} {Fundamental {{Symmetries}} in the {{Era}} of the {{LHC}}},\ \textbf {\bibinfo {volume} {71}},\ \bibinfo {pages} {119} (\bibinfo {year} {2013})}\BibitemShut {NoStop}%
\bibitem [{\citenamefont {Blunden}\ \emph {et~al.}(2012)\citenamefont {Blunden}, \citenamefont {Melnitchouk},\ and\ \citenamefont {Thomas}}]{blundenGammaZBoxCorrections2012}%
  \BibitemOpen
  \bibfield  {author} {\bibinfo {author} {\bibfnamefont {P.~G.}\ \bibnamefont {Blunden}}, \bibinfo {author} {\bibfnamefont {W.}~\bibnamefont {Melnitchouk}},\ and\ \bibinfo {author} {\bibfnamefont {A.~W.}\ \bibnamefont {Thomas}},\ }\bibfield  {title} {\bibinfo {title} {{{gammaZ Box Corrections}} to {{Weak Charges}} of {{Heavy Nuclei}} in {{Atomic Parity Violation}}},\ }\href {https://doi.org/10.1103/PhysRevLett.109.262301} {\bibfield  {journal} {\bibinfo  {journal} {Physical Review Letters}\ }\textbf {\bibinfo {volume} {109}},\ \bibinfo {pages} {262301} (\bibinfo {year} {2012})}\BibitemShut {NoStop}%
\bibitem [{\citenamefont {Craik}(2025)}]{craikEntanglementProtocolMeasure2025}%
  \BibitemOpen
  \bibfield  {author} {\bibinfo {author} {\bibfnamefont {D.~P. L.~A.}\ \bibnamefont {Craik}},\ }\href {https://doi.org/10.48550/arXiv.2503.20003} {\bibinfo {title} {An entanglement protocol to measure atomic parity violation at sub 0.1\% precision}} (\bibinfo {year} {2025}),\ \Eprint {https://arxiv.org/abs/2503.20003} {arXiv:2503.20003 [quant-ph]} \BibitemShut {NoStop}%
\bibitem [{\citenamefont {Viatkina}\ \emph {et~al.}(2019)\citenamefont {Viatkina}, \citenamefont {Antypas}, \citenamefont {Kozlov}, \citenamefont {Budker},\ and\ \citenamefont {Flambaum}}]{viatkinaDependenceAtomicParityviolation2019}%
  \BibitemOpen
  \bibfield  {author} {\bibinfo {author} {\bibfnamefont {A.~V.}\ \bibnamefont {Viatkina}}, \bibinfo {author} {\bibfnamefont {D.}~\bibnamefont {Antypas}}, \bibinfo {author} {\bibfnamefont {M.~G.}\ \bibnamefont {Kozlov}}, \bibinfo {author} {\bibfnamefont {D.}~\bibnamefont {Budker}},\ and\ \bibinfo {author} {\bibfnamefont {V.~V.}\ \bibnamefont {Flambaum}},\ }\bibfield  {title} {\bibinfo {title} {Dependence of atomic parity-violation effects on neutron skins and new physics},\ }\href {https://doi.org/10.1103/PhysRevC.100.034318} {\bibfield  {journal} {\bibinfo  {journal} {Physical Review C}\ }\textbf {\bibinfo {volume} {100}},\ \bibinfo {pages} {034318} (\bibinfo {year} {2019})}\BibitemShut {NoStop}%
\bibitem [{\citenamefont {Emrich}\ \emph {et~al.}(1983)\citenamefont {Emrich}, \citenamefont {Fricke}, \citenamefont {Mallot}, \citenamefont {Miska}, \citenamefont {Sieberling}, \citenamefont {Cavedon}, \citenamefont {Frois},\ and\ \citenamefont {Goutte}}]{emrichRadialDistributionNucleons1983}%
  \BibitemOpen
  \bibfield  {author} {\bibinfo {author} {\bibfnamefont {H.~J.}\ \bibnamefont {Emrich}}, \bibinfo {author} {\bibfnamefont {G.}~\bibnamefont {Fricke}}, \bibinfo {author} {\bibfnamefont {G.}~\bibnamefont {Mallot}}, \bibinfo {author} {\bibfnamefont {H.}~\bibnamefont {Miska}}, \bibinfo {author} {\bibfnamefont {H.~G.}\ \bibnamefont {Sieberling}}, \bibinfo {author} {\bibfnamefont {J.~M.}\ \bibnamefont {Cavedon}}, \bibinfo {author} {\bibfnamefont {B.}~\bibnamefont {Frois}},\ and\ \bibinfo {author} {\bibfnamefont {D.}~\bibnamefont {Goutte}},\ }\bibfield  {title} {\bibinfo {title} {Radial distribution of nucleons in the isotopes 48,{{40Ca}}},\ }\href {https://doi.org/10.1016/0375-9474(83)90034-9} {\bibfield  {journal} {\bibinfo  {journal} {Nuclear Physics A}\ }\textbf {\bibinfo {volume} {396}},\ \bibinfo {pages} {401} (\bibinfo {year} {1983})}\BibitemShut {NoStop}%
\bibitem [{\citenamefont {Yerokhin}\ and\ \citenamefont {Ohayon}(2026)}]{yerokhin2025model}%
  \BibitemOpen
  \bibfield  {author} {\bibinfo {author} {\bibfnamefont {V.~A.}\ \bibnamefont {Yerokhin}}\ and\ \bibinfo {author} {\bibfnamefont {B.}~\bibnamefont {Ohayon}},\ }\bibfield  {title} {\bibinfo {title} {Model-independent determination of nuclear charge radii from {{Li-like}} ions},\ }\href {https://doi.org/10.1103/71d6-w383} {\bibfield  {journal} {\bibinfo  {journal} {Physical Review A}\ }\textbf {\bibinfo {volume} {113}},\ \bibinfo {pages} {012804} (\bibinfo {year} {2026})}\BibitemShut {NoStop}%
\bibitem [{\citenamefont {Ohayon}(2025)}]{OHAYON2025101732}%
  \BibitemOpen
  \bibfield  {author} {\bibinfo {author} {\bibfnamefont {B.}~\bibnamefont {Ohayon}},\ }\bibfield  {title} {\bibinfo {title} {Critical evaluation of reference charge radii and applications in mirror nuclei},\ }\href {https://doi.org/https://doi.org/10.1016/j.adt.2025.101732} {\bibfield  {journal} {\bibinfo  {journal} {Atomic Data and Nuclear Data Tables}\ }\textbf {\bibinfo {volume} {165}},\ \bibinfo {pages} {101732} (\bibinfo {year} {2025})}\BibitemShut {NoStop}%
\bibitem [{\citenamefont {Fricke}\ \emph {et~al.}(2004)\citenamefont {Fricke}, \citenamefont {Heilig},\ and\ \citenamefont {Schopper}}]{fricke2004nuclear}%
  \BibitemOpen
  \bibfield  {author} {\bibinfo {author} {\bibfnamefont {G.}~\bibnamefont {Fricke}}, \bibinfo {author} {\bibfnamefont {K.}~\bibnamefont {Heilig}},\ and\ \bibinfo {author} {\bibfnamefont {H.~F.}\ \bibnamefont {Schopper}},\ }\href@noop {} {\emph {\bibinfo {title} {Nuclear charge radii}}},\ Vol.\ \bibinfo {volume} {454}\ (\bibinfo  {publisher} {Springer Berlin},\ \bibinfo {year} {2004})\BibitemShut {NoStop}%
\bibitem [{\citenamefont {De~Vries}\ \emph {et~al.}(1987)\citenamefont {De~Vries}, \citenamefont {De~Jager},\ and\ \citenamefont {De~Vries}}]{devriesNuclearChargedensitydistributionParameters1987}%
  \BibitemOpen
  \bibfield  {author} {\bibinfo {author} {\bibfnamefont {H.}~\bibnamefont {De~Vries}}, \bibinfo {author} {\bibfnamefont {C.~W.}\ \bibnamefont {De~Jager}},\ and\ \bibinfo {author} {\bibfnamefont {C.}~\bibnamefont {De~Vries}},\ }\bibfield  {title} {\bibinfo {title} {Nuclear charge-density-distribution parameters from elastic electron scattering},\ }\href {https://doi.org/10.1016/0092-640X(87)90013-1} {\bibfield  {journal} {\bibinfo  {journal} {Atomic Data and Nuclear Data Tables}\ }\textbf {\bibinfo {volume} {36}},\ \bibinfo {pages} {495} (\bibinfo {year} {1987})}\BibitemShut {NoStop}%
\bibitem [{\citenamefont {Fricke}\ \emph {et~al.}(1995)\citenamefont {Fricke}, \citenamefont {Bernhardt}, \citenamefont {Heilig}, \citenamefont {Schaller}, \citenamefont {Schellenberg}, \citenamefont {Shera},\ and\ \citenamefont {Dejager}}]{FRICKE1995177}%
  \BibitemOpen
  \bibfield  {author} {\bibinfo {author} {\bibfnamefont {G.}~\bibnamefont {Fricke}}, \bibinfo {author} {\bibfnamefont {C.}~\bibnamefont {Bernhardt}}, \bibinfo {author} {\bibfnamefont {K.}~\bibnamefont {Heilig}}, \bibinfo {author} {\bibfnamefont {L.}~\bibnamefont {Schaller}}, \bibinfo {author} {\bibfnamefont {L.}~\bibnamefont {Schellenberg}}, \bibinfo {author} {\bibfnamefont {E.}~\bibnamefont {Shera}},\ and\ \bibinfo {author} {\bibfnamefont {C.}~\bibnamefont {Dejager}},\ }\bibfield  {title} {\bibinfo {title} {Nuclear ground state charge radii from electromagnetic interactions},\ }\href {https://doi.org/https://doi.org/10.1006/adnd.1995.1007} {\bibfield  {journal} {\bibinfo  {journal} {Atomic Data and Nuclear Data Tables}\ }\textbf {\bibinfo {volume} {60}},\ \bibinfo {pages} {177} (\bibinfo {year} {1995})}\BibitemShut {NoStop}%
\bibitem [{\citenamefont {Sick}\ \emph {et~al.}(1979)\citenamefont {Sick}, \citenamefont {Bellicard}, \citenamefont {Cavedon}, \citenamefont {Frois}, \citenamefont {Huet}, \citenamefont {Leconte}, \citenamefont {Ho},\ and\ \citenamefont {Platchkov}}]{sickChargeDensity40Ca1979}%
  \BibitemOpen
  \bibfield  {author} {\bibinfo {author} {\bibfnamefont {I.}~\bibnamefont {Sick}}, \bibinfo {author} {\bibfnamefont {J.~B.}\ \bibnamefont {Bellicard}}, \bibinfo {author} {\bibfnamefont {J.~M.}\ \bibnamefont {Cavedon}}, \bibinfo {author} {\bibfnamefont {B.}~\bibnamefont {Frois}}, \bibinfo {author} {\bibfnamefont {M.}~\bibnamefont {Huet}}, \bibinfo {author} {\bibfnamefont {P.}~\bibnamefont {Leconte}}, \bibinfo {author} {\bibfnamefont {P.~X.}\ \bibnamefont {Ho}},\ and\ \bibinfo {author} {\bibfnamefont {S.}~\bibnamefont {Platchkov}},\ }\bibfield  {title} {\bibinfo {title} {Charge density of {{40Ca}}},\ }\href {https://doi.org/10.1016/0370-2693(79)90458-1} {\bibfield  {journal} {\bibinfo  {journal} {Physics Letters B}\ }\textbf {\bibinfo {volume} {88}},\ \bibinfo {pages} {245} (\bibinfo {year} {1979})}\BibitemShut {NoStop}%
\bibitem [{\citenamefont {Euteneuer}\ \emph {et~al.}(1978)\citenamefont {Euteneuer}, \citenamefont {Friedrich},\ and\ \citenamefont {Voegler}}]{EUTENEUER1978452}%
  \BibitemOpen
  \bibfield  {author} {\bibinfo {author} {\bibfnamefont {H.}~\bibnamefont {Euteneuer}}, \bibinfo {author} {\bibfnamefont {J.}~\bibnamefont {Friedrich}},\ and\ \bibinfo {author} {\bibfnamefont {N.}~\bibnamefont {Voegler}},\ }\bibfield  {title} {\bibinfo {title} {The charge-distribution differences of 209bi, 208, 207, 206, 204pb and 205, 203tl investigated by elastic electron scattering and muonic x-ray data},\ }\href {https://doi.org/https://doi.org/10.1016/0375-9474(78)90143-4} {\bibfield  {journal} {\bibinfo  {journal} {Nuclear Physics A}\ }\textbf {\bibinfo {volume} {298}},\ \bibinfo {pages} {452} (\bibinfo {year} {1978})}\BibitemShut {NoStop}%
\bibitem [{\citenamefont {Yerokhin}\ and\ \citenamefont {Shabaev}(2015)}]{yerokhinLambShiftStates2015}%
  \BibitemOpen
  \bibfield  {author} {\bibinfo {author} {\bibfnamefont {V.~A.}\ \bibnamefont {Yerokhin}}\ and\ \bibinfo {author} {\bibfnamefont {V.~M.}\ \bibnamefont {Shabaev}},\ }\bibfield  {title} {\bibinfo {title} {Lamb {{Shift}} of n = 1 and n = 2 {{States}} of {{Hydrogen-like Atoms}}, 1 {$\leq$} {{Z}} {$\leq$} 110},\ }\href {https://doi.org/10.1063/1.4927487} {\bibfield  {journal} {\bibinfo  {journal} {Journal of Physical and Chemical Reference Data}\ }\textbf {\bibinfo {volume} {44}},\ \bibinfo {pages} {033103} (\bibinfo {year} {2015})}\BibitemShut {NoStop}%
\bibitem [{\citenamefont {Mertens}()}]{mertensNumericalSolutionDirac}%
  \BibitemOpen
  \bibfield  {author} {\bibinfo {author} {\bibfnamefont {C.~G.}\ \bibnamefont {Mertens}},\ }\bibfield  {title} {\bibinfo {title} {Numerical {{Solution}} of the {{Dirac Equation}} for an {{Arbitrary Central Potential}}},\ }\href@noop {} {\ }\bibinfo {note} {\textit{Bachelor thesis}, {\tt{https://github.com/suppetia/qm-dish}}}\BibitemShut {NoStop}%
\bibitem [{\citenamefont {Yerokhin}\ and\ \citenamefont {Surzhykov}(2018)}]{yerokhinEnergyLevelsCoreExcited2018}%
  \BibitemOpen
  \bibfield  {author} {\bibinfo {author} {\bibfnamefont {V.~A.}\ \bibnamefont {Yerokhin}}\ and\ \bibinfo {author} {\bibfnamefont {A.}~\bibnamefont {Surzhykov}},\ }\bibfield  {title} {\bibinfo {title} {Energy {{Levels}} of {{Core-Excited}} 1s2l2l{$\prime$} {{States}} in {{Lithium-Like Ions}}: {{Argon}} to {{Uranium}}},\ }\href {https://doi.org/10.1063/1.5034574} {\bibfield  {journal} {\bibinfo  {journal} {Journal of Physical and Chemical Reference Data}\ }\textbf {\bibinfo {volume} {47}},\ \bibinfo {pages} {023105} (\bibinfo {year} {2018})}\BibitemShut {NoStop}%
\bibitem [{\citenamefont {J{\"o}nsson}\ \emph {et~al.}(2007)\citenamefont {J{\"o}nsson}, \citenamefont {He}, \citenamefont {Froese~Fischer},\ and\ \citenamefont {Grant}}]{jonssonGrasp2KRelativisticAtomic2007}%
  \BibitemOpen
  \bibfield  {author} {\bibinfo {author} {\bibfnamefont {P.}~\bibnamefont {J{\"o}nsson}}, \bibinfo {author} {\bibfnamefont {X.}~\bibnamefont {He}}, \bibinfo {author} {\bibfnamefont {C.}~\bibnamefont {Froese~Fischer}},\ and\ \bibinfo {author} {\bibfnamefont {I.~P.}\ \bibnamefont {Grant}},\ }\bibfield  {title} {\bibinfo {title} {The {{grasp2K}} relativistic atomic structure package},\ }\href {https://doi.org/10.1016/j.cpc.2007.06.002} {\bibfield  {journal} {\bibinfo  {journal} {Computer Physics Communications}\ }\textbf {\bibinfo {volume} {177}},\ \bibinfo {pages} {597} (\bibinfo {year} {2007})}\BibitemShut {NoStop}%
\bibitem [{\citenamefont {Kramida}\ \emph {et~al.}(2024)\citenamefont {Kramida}, \citenamefont {{Yu.~Ralchenko}}, \citenamefont {Reader},\ and\ \citenamefont {{and NIST ASD Team}}}]{nist}%
  \BibitemOpen
  \bibfield  {author} {\bibinfo {author} {\bibfnamefont {A.}~\bibnamefont {Kramida}}, \bibinfo {author} {\bibnamefont {{Yu.~Ralchenko}}}, \bibinfo {author} {\bibfnamefont {J.}~\bibnamefont {Reader}},\ and\ \bibinfo {author} {\bibnamefont {{and NIST ASD Team}}},\ }\href@noop {} {}\bibinfo {howpublished} {{NIST Atomic Spectra Database (ver. 5.12), [Online]. Available: {\tt{https://physics.nist.gov/asd}} [2025, June 12]. National Institute of Standards and Technology, Gaithersburg, MD.}} (\bibinfo {year} {2024})\BibitemShut {NoStop}%
\bibitem [{\citenamefont {Kahl}\ and\ \citenamefont {Berengut}(2019)}]{kahlAMBiTProgrammeHighprecision2019}%
  \BibitemOpen
  \bibfield  {author} {\bibinfo {author} {\bibfnamefont {E.~V.}\ \bibnamefont {Kahl}}\ and\ \bibinfo {author} {\bibfnamefont {J.~C.}\ \bibnamefont {Berengut}},\ }\bibfield  {title} {\bibinfo {title} {{{AMBiT}}: {{A}} programme for high-precision relativistic atomic structure calculations},\ }\href {https://doi.org/10.1016/j.cpc.2018.12.014} {\bibfield  {journal} {\bibinfo  {journal} {Computer Physics Communications}\ }\textbf {\bibinfo {volume} {238}},\ \bibinfo {pages} {232} (\bibinfo {year} {2019})}\BibitemShut {NoStop}%
\bibitem [{\citenamefont {Kozlov}\ \emph {et~al.}(2025)\citenamefont {Kozlov}, \citenamefont {Kaygorodov}, \citenamefont {Demidov},\ and\ \citenamefont {Yerokhin}}]{kozlovSelfenergyCorrectionE12025}%
  \BibitemOpen
  \bibfield  {author} {\bibinfo {author} {\bibfnamefont {M.~G.}\ \bibnamefont {Kozlov}}, \bibinfo {author} {\bibfnamefont {M.~Y.}\ \bibnamefont {Kaygorodov}}, \bibinfo {author} {\bibfnamefont {{\relax Yu}.~A.}\ \bibnamefont {Demidov}},\ and\ \bibinfo {author} {\bibfnamefont {V.~A.}\ \bibnamefont {Yerokhin}},\ }\bibfield  {title} {\bibinfo {title} {Self-energy correction to the e1 transition amplitudes in hydrogenlike ions},\ }\href {https://doi.org/10.1103/PhysRevA.111.022816} {\bibfield  {journal} {\bibinfo  {journal} {Physical Review A}\ }\textbf {\bibinfo {volume} {111}},\ \bibinfo {pages} {022816} (\bibinfo {year} {2025})}\BibitemShut {NoStop}%
\bibitem [{\citenamefont {Bouchiat}\ and\ \citenamefont {Bouchiat}(1974)}]{bouchiatParityViolationInduced1974}%
  \BibitemOpen
  \bibfield  {author} {\bibinfo {author} {\bibfnamefont {M.~A.}\ \bibnamefont {Bouchiat}}\ and\ \bibinfo {author} {\bibfnamefont {C.}~\bibnamefont {Bouchiat}},\ }\bibfield  {title} {\bibinfo {title} {I. {{Parity}} violation induced by weak neutral currents in atomic physics},\ }\href {https://doi.org/10.1051/jphys:019740035012089900} {\bibfield  {journal} {\bibinfo  {journal} {Journal de Physique}\ }\textbf {\bibinfo {volume} {35}},\ \bibinfo {pages} {899} (\bibinfo {year} {1974})}\BibitemShut {NoStop}%
  \bibitem [{\citenamefont {Ferro}\ \emph {et~al.}(2011)\citenamefont {Ferro}, \citenamefont {Surzhykov},\ and\ \citenamefont {St{\"o}hlker}}]{ferroHyperfineTransitionsHelike2011}%
    \BibitemOpen
    \bibfield  {author} {\bibinfo {author} {\bibfnamefont {F.}~\bibnamefont {Ferro}}, \bibinfo {author} {\bibfnamefont {A.}~\bibnamefont {Surzhykov}},\ and\ \bibinfo {author} {\bibfnamefont {T.}~\bibnamefont {St{\"o}hlker}},\ }\bibfield  {title} {\bibinfo {title} {Hyperfine transitions in {{He-like}} ions as a tool for nuclear-spin-dependent parity-nonconservation studies},\ }\href {https://doi.org/10.1103/PhysRevA.83.052518} {\bibfield  {journal} {\bibinfo  {journal} {Physical Review A}\ }\textbf {\bibinfo {volume} {83}},\ \bibinfo {pages} {052518} (\bibinfo {year} {2011})}\BibitemShut {NoStop}%
  \bibitem [{\citenamefont {Flambaum}\ \emph {et~al.}(2017)\citenamefont {Flambaum}, \citenamefont {Dzuba},\ and\ \citenamefont {Harabati}}]{flambaumEffectNuclearQuadrupole2017}%
    \BibitemOpen
    \bibfield  {author} {\bibinfo {author} {\bibfnamefont {V.~V.}\ \bibnamefont {Flambaum}}, \bibinfo {author} {\bibfnamefont {V.~A.}\ \bibnamefont {Dzuba}},\ and\ \bibinfo {author} {\bibfnamefont {C.}~\bibnamefont {Harabati}},\ }\bibfield  {title} {\bibinfo {title} {Effect of nuclear quadrupole moments on parity nonconservation in atoms},\ }\href {https://doi.org/10.1103/PhysRevA.96.012516} {\bibfield  {journal} {\bibinfo  {journal} {Physical Review A}\ }\textbf {\bibinfo {volume} {96}},\ \bibinfo {pages} {012516} (\bibinfo {year} {2017})}\BibitemShut {NoStop}%
\end{thebibliography}

%

\end{document}